\documentclass[12pt,english]{article}
\usepackage[margin=2.54cm]{geometry} %rand
\usepackage{setspace}
\usepackage{graphicx}

\usepackage{fixltx2e}
\usepackage{amssymb}
\usepackage[T1]{fontenc}
\usepackage[latin9]{inputenc}
\usepackage{textgreek}
\usepackage{babel}

\date{} %kein datum

\title{Giant magnetic anisotropy and tunnelling of the magnetization in Li$_2$(Li$_{1-x}$Fe$_x$)N
}
\begin{document}
\maketitle
% \vspace{-8ex}
\noindent
\author{A. Jesche$^{1,*}$, 
R. W. McCallum$^{1,2}$, 
S. Thimmaiah$^1$,
J. L. Jacobs$^{1,3}$,
V. Taufour$^4$,
A.~Kreyssig$^{1,4}$,
R. S. Houk$^{1,3}$,
S.~L.~Bud'ko$^{1,4}$ \& P. C. Canfield$^{1,4}$}
\\
$^1$\textit{The Ames Laboratory, Iowa State University, Ames, Iowa, USA}\\
$^2$\textit{Department of Materials Science and Engineering, Iowa State University, Ames, Iowa, USA}\\
$^3$\textit{Department of Chemistry, Iowa State University, Ames, Iowa, USA}\\
$^4$\textit{Department of Physics and Astronomy, Iowa State University, Ames, Iowa, USA}\\
$^\star$jesche@ameslab.gov

\begin{abstract}
\noindent 
Large magnetic anisotropy and coercivity are key properties of functional magnetic materials and are generally associated with rare earth elements. 
Here we show an extreme, uniaxial magnetic anisotropy and the emergence of magnetic hysteresis in Li\textsubscript{2}(Li\textsubscript{1-\textit{x}}Fe\textsubscript{\textit{x}})N.
An extrapolated, magnetic anisotropy field of 220\,Tesla and a coercivity field of over 11 Tesla at 2 Kelvin outperform all known hard-ferromagnets and single-molecule magnets (SMMs). 
Steps in the hysteresis loops and relaxation phenomena in striking similarity to SMMs are particularly pronounced for \textit{x}\,$\ll$\,1 
and indicate the presence of nano-scale magnetic centres.
Quantum tunnelling, in form of temperature-independent relaxation and coercivity, deviation from Arrhenius behaviour and blocking of the relaxation, dominates the magnetic properties up to 10 Kelvin. 
The simple crystal structure, the availability of large single crystals, and the ability to vary the Fe concentration make Li\textsubscript{2}(Li\textsubscript{1-\textit{x}}Fe\textsubscript{\textit{x}})N (i) an ideal model system to study macroscopic quantum effects at elevated temperatures and (ii) a basis for novel functional magnetic materials.
\end{abstract}

\section*{Introduction}
Controlling individual spins on an atomic level is one of the major goals of solid state physics and chemistry. 
To this effect, single-molecule magnets (SMMs)\,\cite{Sessoli2003} have brought significant insight ranging from fundamental quantum effects like tunnelling of the magnetization\,\cite{Paulsen1995} and quantum decoherence\,\cite{Takahashi2011} to possible applications in quantum computing\,\cite{Leuenberger2001} and high density data storage\,\cite{Bogani2008}. 
Their basic magnetic units are coupled spins of a few magnetic atoms which are embedded in and separated by complex organic structures.  
Whereas these are small magnetic units, they are still finite in extent and need to carefully balance coupling between magnetic atoms and isolation of one molecule from the next. 
Dimers of transition metals, as the small-size end-point of SMMs, have been theoretically proposed to be promising candidates for novel information storage devices\,\cite{Strandberg2007}.
An alternative approach for the design of magnetic materials based on a few or even single atoms as magnetic units are ad-atoms on metallic surfaces\,\cite{Gambardella2003, Khajetoorians2013}. 
The key property among these actually very different examples, and of any nano-scale magnetic system, is a large magnetic anisotropy energy. 

Basic magnetic units of SMMs are transition metal ion clusters\,\cite{Sessoli1993}, lanthanide ion clusters\,\cite{Lin2009,Rinehart2011}, or mixed clusters of both\,\cite{Osa2004}.
Even mononuclear complexes based on a single lanthanide ion have been realized\,\cite{Ishikawa2003, Jiang2010}. 
The lanthanide based systems are promising due to their large single ion anisotropy\,\cite{Ishikawa2007,Sessoli2012} which often leads to large magnetic anisotropy energies.
In contrast, single transition metal ions are seldom considered as suitable candidates and the number of reported attempts to use them as mononuclear magnetic units is limited\,\cite{Freedman2010, Zadrozny2011, Zadrozny2013p}.
The main reason is the widely known paradigm of 'orbital quenching'. 
This suppression of the orbital contribution to the magnetic moment by the crystal electric field leads to 
a comparatively small anisotropy energy (neglecting spin-orbit coupling, a pure spin contribution is by default isotropic). 
The absence of an orbital contribution is reflected, e.g., in the largely isotropic magnetization of the elemental ferromagnets Fe, Co, and Ni\,\cite{Honda1926, Kaya1928}.

However, Klatyk \textit{et\,al.}\,\cite{Klatyk2002}~have suggested that a rare interplay of crystal electric field effects and spin-orbit coupling causes a large orbital contribution to the magnetic moment of Fe in polycrystalline Li$_2$(Li$_{1-x}$Fe$_{x}$)N.
Based on the strong increase of the magnetization upon cooling\,\cite{Klatyk2002} and on M\"ossbauer spectroscopy\,\cite{Klatyk2002,Ksenofontov2003} a ferromagnetic ordering with $T_{\rm C} \approx 65$\,K was inferred for $x = 0.21$ and, furthermore, huge hyperfine fields were found.
The orbital contribution to the magnetic moment of Fe as well as the large hyperfine fields were theoretically described within the framework of local density approximation (LDA) calculations\,\cite{Klatyk2002,Novak2002}. 
Furthermore, a large magnetic anisotropy has been theoretically proposed\,\cite{Klatyk2002,Novak2002}. 

Here we show the experimental verification of the large anisotropy by magnetization measurements on single crystals and reveal a huge magnetic hysteresis as the key property of functional magnetic materials. 
More importantly, we have discovered that this highly anisotropic transition metal system manifests strong indications of a macroscopic quantum tunnelling of the magnetization in the form of pronounced steps in the magnetization loops and a temperature-independent relaxation.
Tunnelling of the magnetization explicitly refers to the macroscopic tunnelling of the total magnetization and not to microscopic tunnelling events influencing the domain-wall movement in ferromagnets\,\cite{Uehara1986}. 

\begin{figure}
\center
\includegraphics[width=0.8\textwidth]{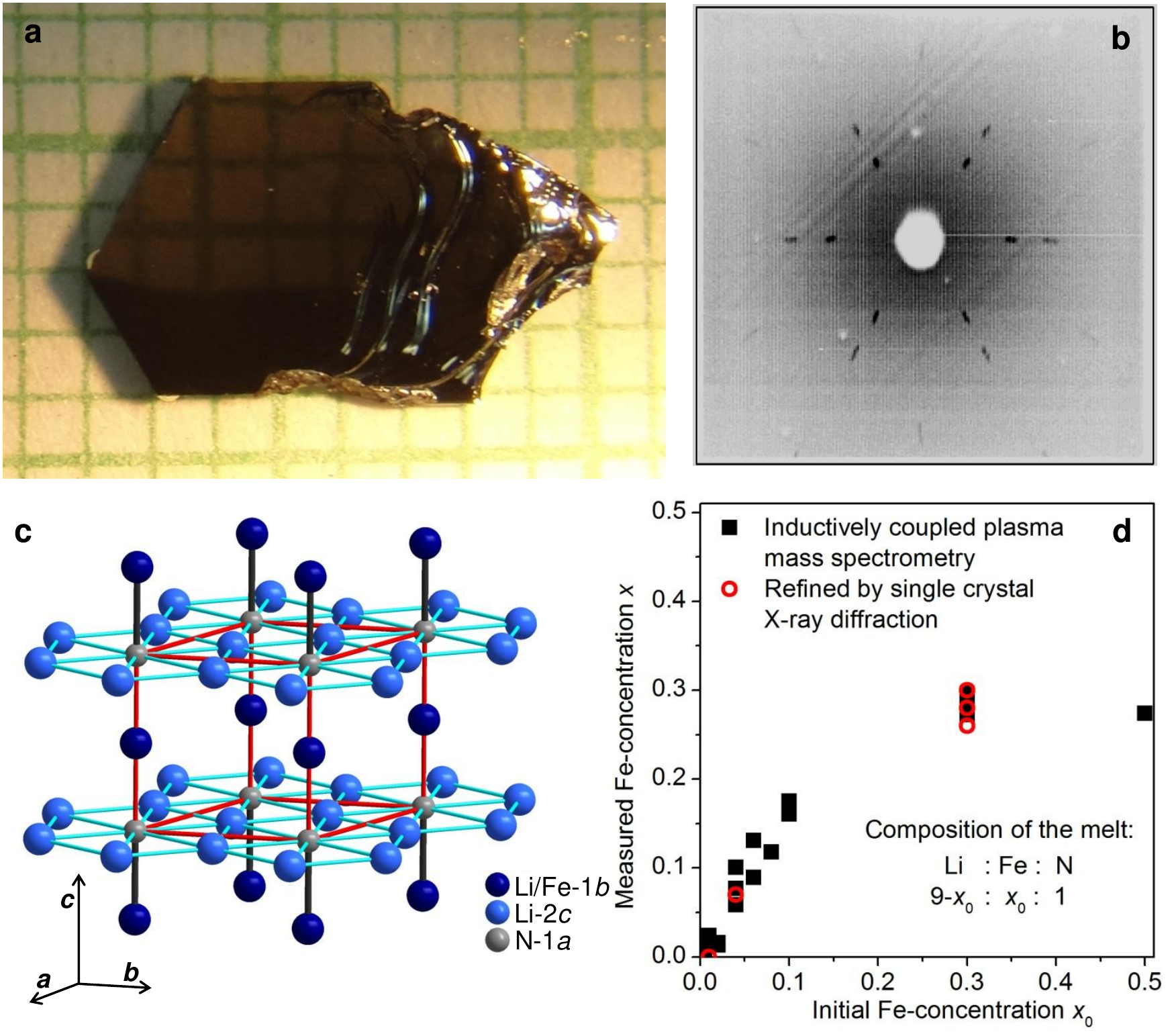}
\caption{{Basic structural properties}. \textbf{a,} Single crystal of Li$_2$(Li$_{0.90}$Fe$_{0.10}$)N on a millimetre grid and \textbf{b,} corresponding Laue-back-reflection pattern. The crystal is not transparent and the faint grid pattern is a reflection off of the lens and the flat surface of the reflecting top facet. \textbf{c,} Crystal structure with Li$_2$N layers separated by a second Li-site which is partially occupied by Fe. The unit cell of the hexagonal lattice is indicated by red lines.
\textbf{d,} Measured Fe concentration \textit{x} as a function of the Fe concentration in the melt \textit{x}$_0$. For small concentrations, \textit{x} tends to be larger than \textit{x}$_0$, however, a plateau in \textit{x} as a function of \textit{x}$_0$ emerges for \textit{x}$_0 \gtrsim$ 0.3. 
As discussed in Supplementary Methods, \textit{x} was determined by both inductively coupled plasma mass spectroscopy and by refinement of single crystal X-ray diffraction data.} 
\label{crystal}
\end{figure}
The Li$_3$N host provides an extremely anisotropic ligand field for the Fe atoms as well as an insulating environment in analogy to the 'organic framework' of SMMs.
There are no indications for meso- or macroscopic phase separation (see SI\,\ref{phase}).
Although we can not completely rule out cluster formation (e.g. dimers or trimers of Fe on adjacent Li sites) the preponderance of the data support a single iron atom as mononuclear magnetic centre which is, furthermore, the simplest model and based on the fewest assumptions.
The phenomenological similarities to SMMs indicate that the spontaneous magnetization and hysteresis are primarily caused by the extreme magnetic anisotropy and not by collective ordering phenomena. 
In accordance with earlier work on mononuclear systems, Li$_2$(Li$_{1-x}$Fe$_x$)N might be considered as 'atomic magnet'\,\cite{Giraud2003} or 'single-ionic SMM'\,\cite{Ishikawa2007}.
The magnetic anisotropy, coercivity and energy barrier for spin-inversion found in Li$_2$(Li$_{1-x}$Fe$_x$)N are roughly one order of magnitude larger than in typical SMMs. 
Magnetic hysteresis exists up to comparatively high temperatures of $T \gtrsim 16$\,K for \textit{x} $\ll 1$ and is further enhanced up to $T \gtrsim 50$\,K for the largest Fe concentration of $x = 0.28$. 

\section*{Results}

\subsection*{Basic properties of Li$_2$(Li$_{1-x}$Fe$_x$)N}

We grew Li$_2$(Li$_{1-x}$Fe$_x$)N single crystals of several millimetre size (Figs.\,\ref{crystal}a,\,b) and Fe concentrations ranging over three orders of magnitude $x = 0.00028$ to $0.28$ by using a Li-flux method to create a rare, nitrogen-bearing metallic solution.
Li$_2$(Li$_{1-x}$Fe$_x$)N crystallizes in a hexagonal lattice, space group $P\,6/m\,m\,m$, with a rather simple unit cell (Fig.\,\ref{crystal}c) and lattice parameters of $a = 3.652(8)$\,\AA\, and $c = 3.870(10)$\,\AA\,for $x = 0$. 
Fe-substitution causes an increase of $a$ but a decrease of $c$ by 1.1\,\% and 1.5\,\%, respectively, for $x = 0.28$\, with intermediate concentrations showing a linear dependence on $x$ following Vegards law (Supplementary Fig.\,1). 
As indicated by the notation of the chemical formula, the substituted Fe-atoms occupy only the Li-1$b$ Wyckoff position which is sandwiched between Li$_2$N layers.
The iron concentrations, $x$, were measured by inductively coupled plasma mass spectrometry which enables a quantitative analysis on a parts per billion level (Fig.\,\ref{crystal}d).
Structural parameters and selected $x$ values were determined by single crystal and powder X-ray diffraction and are in good agreement with earlier results\,\cite{Klatyk2002, Rabenau1976, Klatyk1999, Yamada2011}.
To support our findings we present details of crystal growth procedure and chemical analysis (see Methods), as well as X-ray powder diffraction (Supplementary Fig.\,2, Supplementary Note\,2) and  X-ray single crystal diffraction (Supplementary Tables\,1\,and\,2, Supplementary Note\,3). 

The samples are air sensitive in powder form but visual inspection and magnetic measurements revealed no significant decay of larger single crystals on a timescale of hours. 
As stated in Ref.\,\cite{Gregory2001}, this is probably "due, somewhat perversely, to the formation of a surface film of predominantly LiOH".
Covering the samples with a thin layer of Apiezon M grease further protects the sample and no degradation was observed over a period of several weeks. 
Single crystals which had been exposed to air for a few minutes and were stored afterwards in an inert atmosphere (argon or nitrogen) did not change their magnetic properties on a time scale of three months. 
The electrical resistivity at room temperature is estimated to be $\rho > 10^5\,\Omega$\,cm for all studied $x$. 

\subsection*{Magnetization}

\begin{figure}
\center
\includegraphics[width=0.9\textwidth]{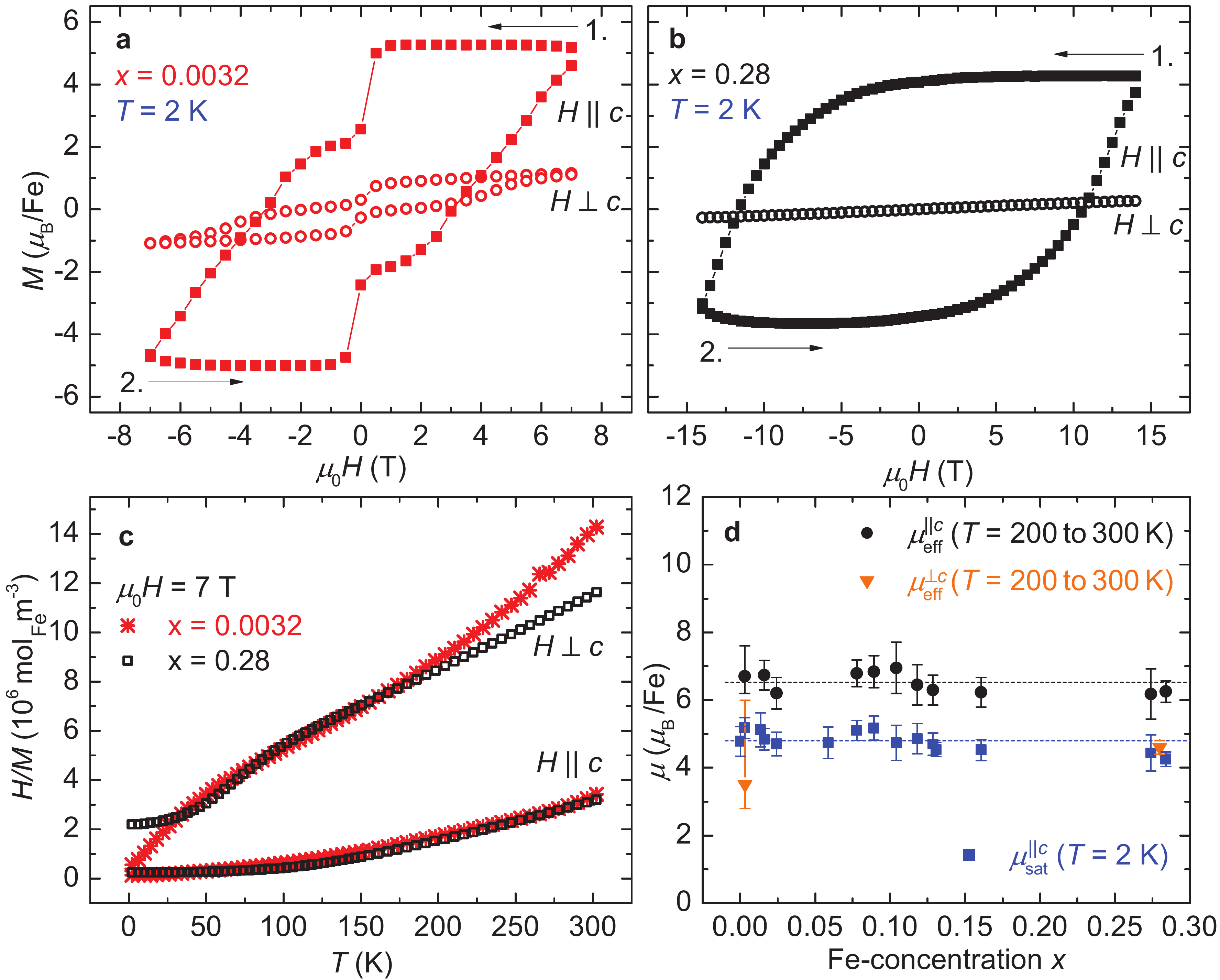}
\caption{Basic magnetic properties of Li$_2$(Li$_{1-x}$Fe$_x$)N.
\textbf{a,\,b,} Large hysteresis in the magnetization, \textit{M(H)}, and a pronounced anisotropy depending on the orientation of the applied magnetic field, \textit{H}, with respect to the crystallographic axes for \textit{x} = 0.0032 and \textit{x} = 0.28. 
Whereas \textit{M} can be saturated for \textit{H} $\parallel$ \textit{c} it is slowly increasing with \textit{H} for \textit{H} $\perp$ \textit{c} up to the highest available fields. 
\textbf{c,} The temperature dependence of the inverse magnetization for \textit{x} = 0.28 follows a Curie-Weiss-law for \textit{T} $\gtrsim$ 150\,K with effective moments of $\mu_{\rm eff}^{\perp c}$ = 4.6(3)\,$\mu_{\rm B}$ for \textit{H} $\perp$ \textit{c} and $\mu_{\rm eff}^{\parallel c}$ = 6.3(4)\,$\mu_{\rm B}$ for \textit{H} $\parallel$ \textit{c}.
Similar behaviour is observed for the two orders of magnitude lower concentration of \textit{x} = 0.0032 where the deviation for \textit{H} $\perp$ \textit{c} at \textit{T} $>$ 200\,K can be caused by a temperature-independent diamagnetic background of the Li$_3$N host which is negligible for higher Fe concentrations.  
\textbf{d,} Both, $\mu_{\rm sat}$ and $\mu_{\rm eff}$ were found to be largely independent of $x$ over the whole investigated range of \textit{x}  = 0.00028 to 0.28. 
The error bars are calculated based on the errors in assessing the sample mass and the Fe concentration and, for $x = 0.0032$, on a diamagnetic contribution of the Li$_3$N host.
The average values are $\mu_{\rm sat}^{\parallel c}$ = 4.8(4) and $\mu_{\rm eff}^{\parallel c}$ = 6.5(4)\,$\mu_{\rm B}$.\label{mag-intro}}
\end{figure}

Extreme magnetic anisotropy of Li$_2$(Li$_{1-x}$Fe$_x$)N is conspicuously evident in the magnetization measurements shown in Figs.\,\ref{mag-intro}a,b for two very different Fe concentrations of \textit{x} = 0.0032 and \textit{x} = 0.28. 
For magnetic field applied along the $c$-axis, $H \parallel c$, the magnetization is constant for $\mu_0H > 1$\,T (starting from the field cooled state) with a large saturated moment of $\mu_{\rm sat}^{\parallel c} \approx 5\,\mu_{\rm B}$ per Fe atom.
In contrast, the magnetization for $H \perp c$ is smaller and slowly increases with field.
A linear extrapolation to higher magnetic fields yields huge anisotropy fields of $\mu_0H_{\rm ani} \approx 88$\,T ($x = 0.0032$) and $\mu_0H_{\rm ani} \approx 220$\,T ($x = 0.28$) defined as the field strength where both magnetization curves intersect. 
The larger magnetic anisotropy field for $x = 0.28$ is reflected in the larger coercivity field of $\mu_0H_{\rm c} = 11.6$\,T found for this concentration. The inverse magnetic susceptibility, $\chi^{-1} = H/M$, roughly follows a Curie-Weiss behaviour (Fig.\,\ref{mag-intro}c) for $T \gtrsim 150$\,K and
 is strongly anisotropic over the whole investigated temperature range.
The corresponding Weiss temperatures are $\Theta_{\rm W}^{\perp c} = -80(10)$\,K and $\Theta_{\rm W}^{\parallel c} = 100(10)$\,K.
In the following we discuss only measurements with $H \parallel c$ since available laboratory fields do not allow saturation of the magnetization for $H \perp c$.
Both $\mu_{\rm sat}^{\parallel c}$ and $\mu_{\rm eff}^{\parallel c}$ were found to be largely independent of the Fe concentration but for a small tendency to decrease with increasing $x$ (Fig.\,\ref{mag-intro}d, see Supplementary Data\,\ref{SI-mag2} for error analysis). 
The average values are $\mu_{\rm sat}^{\parallel c} = 4.8(4)\,\mu_{\rm B}$ and $\mu_{\rm eff}^{\parallel c} = 6.5(4)\,\mu_{\rm B}$ in good agreement with theoretical calculations for $x = 0.17$\,\cite{Novak2002}. 
Furthermore, the obtained effective moment is surprisingly close to the $\mu_{\rm eff} = 6.6\,\mu_{\rm B}$ expectation value of a fully spin-orbit coupled state (Hund's rule coupling) when assuming a $3d^7$ configuration for the proposed Fe$^{+1}$ state (for a discussion of this unusual valence see\,\cite{Klatyk2002, Novak2002}).  

\begin{figure}
\center
\includegraphics[width=0.9\textwidth]{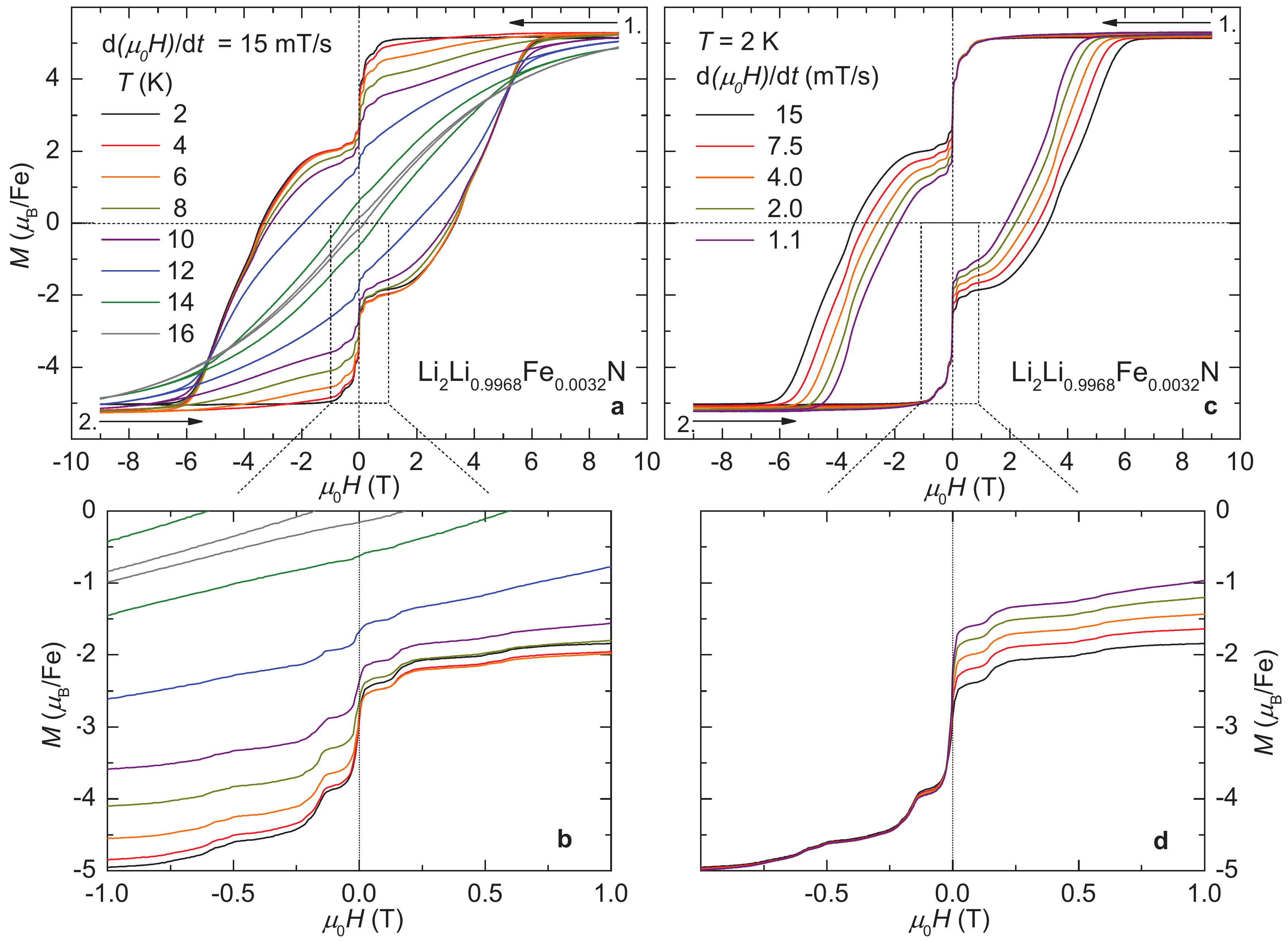}
\caption{Temperature and sweep-rate dependence of the magnetization of Li$_2$(Li$_{1-x}$Fe$_x$)N for \textit{x} = 0.0032, \textit{H} $\parallel$ \textit{c}.
\textbf{a,} Hysteresis emerges for \textit{T} $\leq$ 16\,K and the coercivity fields, \textit{H}$_{\rm c}$, are essentially temperature-independent below \textit{T} = 10\,K. 
\textbf{b,} The pronounced sweep-rate dependence of the magnetization reveals the dynamic nature of the hysteresis.
\textbf{c,} The clear steps in the magnetization are smeared out with increasing temperature and disappear for \textit{T}\,$\gtrsim$\,16\,K.
\textbf{d,} Only the step at \textit{H} = 0 depends significantly on the sweep-rate indicating that the relaxation at the smaller steps at $\mu_0$\textit{H} = $\pm$ 0.15, $\pm$ 0.55\,T is fast on the time-scale of the experiment. 
The step at \textit{H} = 0 is attributed to a flip of the Fe magnetic moment from below to above the crystallographic \textit{a-b} plane.
The energy barrier associated with this transition seems to dominate the relaxation process and forms the global maximum in the magnetic anisotropy energy.
The smaller steps are accordingly associated with tunnelling through smaller, local maxima.  
\label{M-H}}
\end{figure}

In the following we focus on experimental results on the very dilute case which approaches an ideally non-interacting system of single magnetic atoms.
The temperature-dependence of the $M$-$H$ loops for $x = 0.0032$ with a high density of data points are shown in Fig.\,\ref{M-H}a.
Three main observations can be made: 

(i) The magnetization curve is essentially temperature-independent for $T < 10$\,K with $\mu_0H_{\rm c} = 3.4$\,T which indicates the irrelevance of thermal excitations. In contrast, $H_{\rm c}$ changes dramatically between 10 and 16\,K indicating a distinct separation into a low- and a high-temperature behaviour.

(ii) There are pronounced steps in the magnetization. An enlarged view on part of the $M$-$H$ loop, Fig.\,\ref{M-H}c, reveals steps in $M$ at $\mu_0H = 0,\,\pm0.15$, and $\pm0.55$\,T. Another, smaller but well defined, step occurs at $\mu_0H = \pm5$\,T. 
The step-sizes are larger when $H$ approaches zero and decrease with increasing temperature in contrast to the step-positions which are independent of temperature. No steps could be resolved for $T \gtrsim 16$\,K. The steps are strongly suppressed with increasing $x$ (Supplementary Fig.\,\ref{SI-mag}), most likely caused by an increasing Fe-Fe interaction and vanish for $x = 0.28$ (Fig.\,\ref{mag-intro}b).

(iii) The $M$-$H$ loop at $T = 2$\,K starts from a saturated state where the Fe-magnetic moments appear to be aligned parallel to the field, $\mu \parallel H$ with $H \parallel c$ (Fig.\,\ref{M-H}a). The first step occurs already in the first quadrant (upper right corner) when $H$ approaches zero from positive values. 
Similar behaviour was observed in lanthanide-based mononuclear SMMs\,\cite{Ishikawa2007} whereas SMMs based on clusters show no corresponding steps in the first quadrant\,\cite{Sessoli2003}.
A decrease of the magnetization in the first quadrant, where $H$ is still parallel to the initial magnetization, is incompatible with the $c$-axis being the easy axis.
Rather, the moment seems to be tilted away from the $c$-axis and the reduced magnetization is the projection of the moment along the $c$-axis (see Fig.\,\ref{model}a below).
Notice that an axis canted away from the $c$-axis would be 12- or 24-fold degenerate (depending on whether it is oriented along a high-symmetry direction like $h\,0\,l$ or not).
Accordingly, the magnetization perpendicular to the $c$ axis ($H \perp c$, $M \perp c$) is still small in zero field because the perpendicular components of the tilted moments cancel out.
\\
In contrast to a common ferromagnet, the $M$-$H$ loops not only depend on temperature but do also show a pronounced, characteristic dependence on the sweep-rate of the applied magnetic field. Figures\,\ref{M-H}b,d show $M$-$H$ loops at $T = 2$\,K for sweep rates between 15\,mT/s and 1.1\,mT/s, corresponding to a total time between 40 minutes and 9 hours and 10 minutes, respectively, for the whole loop.
The step-sizes at $\mu_0H = \pm0.15$ and $\pm0.55$\,T hardly depend on the sweep-rate, whereas the step-size at $H = 0$ does. 
A low sweep-rate dependence of the step-size is in accordance with a large tunnelling gap (which is much smaller than the energy gap), see\,\cite{Wernsdorfer1999} and references therein.
Although the field values for the steps do not change significantly with sweep-rate, the width of the whole hysteresis loop (and therefore $H_{\rm c}$) does (Fig.\,\ref{M-H}b).
Pronounced sweep-rate dependence of $H_{\rm c}$ is observed also for higher Fe concentrations - see Supplementary Fig.\,\ref{SI-mag}.
The observation of smaller $H_{\rm c}$ values in slower measurements clearly reveals the dynamic nature of the hysteresis. 
This observation motivated a detailed study of the time dependence of the magnetization, which is presented in the following section.

\subsection*{Relaxation}
\begin{figure}
\center
\includegraphics[width=0.94\textwidth]{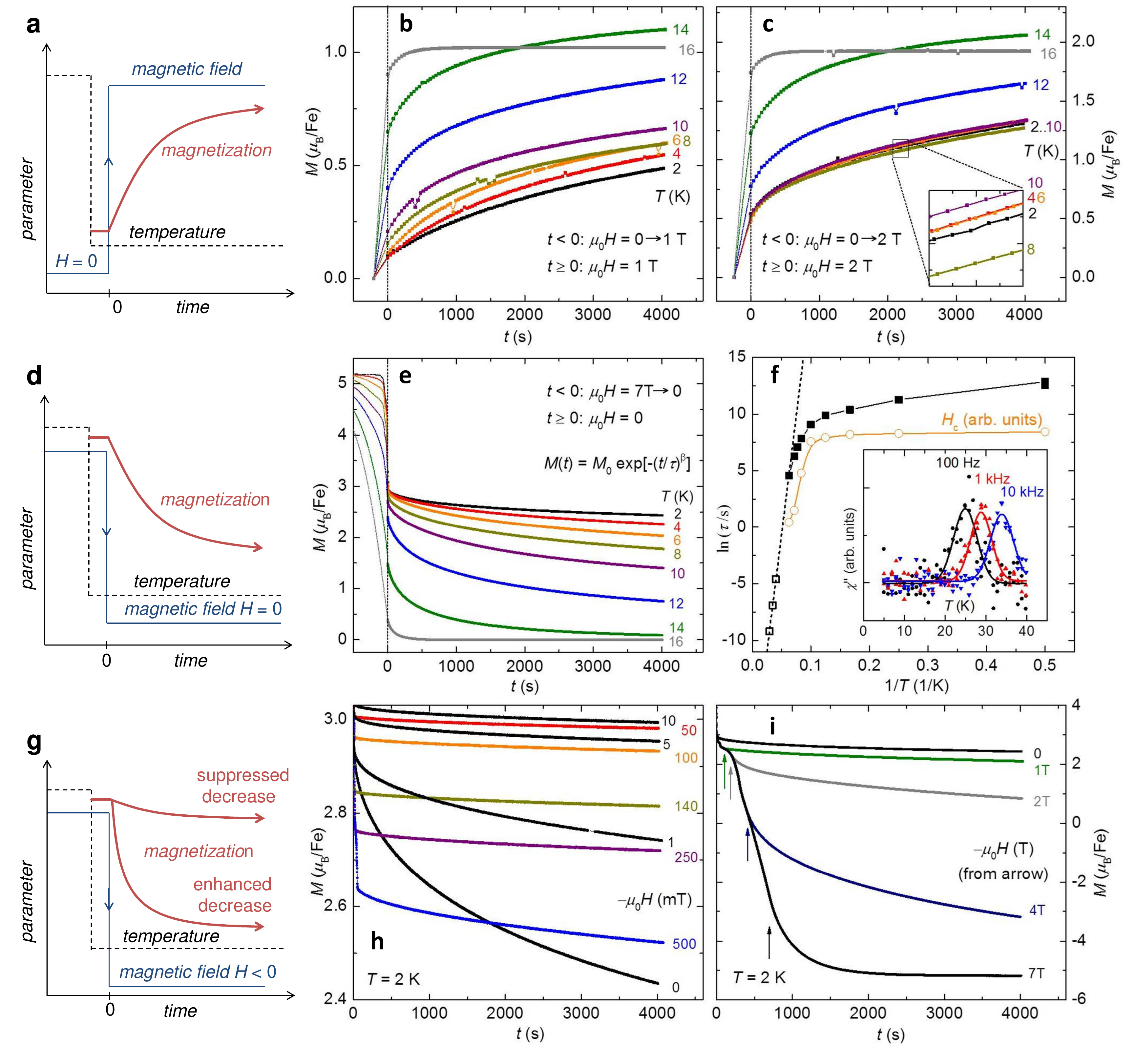}
\caption{{Time-dependence of the magnetization.} 
\textbf{a,\,d,\, and g,} Schematics of different relaxation processes and corresponding experimental data on Li$_2$(Li$_{0.9968}$Fe$_{0.0032}$)N to the right. 
\textbf{b,} The increase in the magnetization as a function of time, \textit{M(t)}, below \textit{T} = 10\,K is only weakly temperature dependent for $\mu_0$\textit{H} = 1\,T.
\textbf{c,} Increasing the applied field to $\mu_0$\textit{H} = 2\,T leads to an essentially temperature-independent relaxation below \textit{T} = 10\,K which is inconsistent with a thermally activated relaxation process. 
\textbf{e,} Decrease in \textit{M(t)} after ramping the field from $\mu_0$\textit{H} = 7\,T to 0. \textit{M(t)} is fit to a stretched exponential function and the obtained relaxation times, $\tau$, are shown as closed squares in form of an Arrhenius-plot in \textbf{f}. 
\textbf{f,} $\tau$ determined from the imaginary part of the ac-magnetic susceptibility, $\chi$\textit{''(T)} (inset), is shown by open squares. 
Thermally activated behaviour is observed at higher temperatures (dashed line). The formation of a plateau towards lower temperatures is incompatible with thermally activated behaviour and indicates the relevance of quantum tunnelling. The coercivity field (open circles, sweep-rate 15\,mT/s) shows similar temperature-dependence indicating slow relaxation as the origin of spontaneous magnetization. 
\textbf{g,} Small negative fields of a few mT, applied opposite to the initial applied field of $\mu_0$\textit{H} = 7\,T, significantly reduce the decrease in \textit{M(t)} compared to \textit{H} = 0. 
\textbf{i,} In contrast, \textit{M(t)} decreases faster in larger negative fields of $\vert\mu_0$\textit{H}$\vert\geq$ 1\,T.  
\label{relax}
}
\end{figure}

When a magnetic moment is subject to a change in the applied magnetic field it will take a finite relaxation time to reach the equilibrium state. 
A ferromagnet reaches this state typically within a fraction of a second (see e.g. Ref.\cite{Cullity1972}.
The sweep-rate dependence of the hysteresis loops indicates a time-scale of several hours for the relaxation in Li$_2$(Li$_{1-x}$Fe$_x$)N. 
This is so slow that the change of the magnetization with time, $M(t)$, during the relaxation process can be directly measured with standard laboratory magnetometers.
Three procedures are schematically shown in the left panels of Fig.\,\ref{relax} (see Supplementary Data\,\ref{protocol} for detailed measurement protocols). The corresponding experimental data are in the panels to the right:  

{\bf Relaxation after switching on the field - Figs.\,\ref{relax}\,a-c.}
$M(t)$ increases rapidly for $T = 16$\,K and the magnetization is constant after $t \sim 500$\,s.
The relaxation is becoming markedly slower as the measurement temperature is reduced to $T = 10$\,K, and $M(t)$ keeps increasing up to the maximum measurement time. 
However, below 10\,K, $M(t)$ is only weakly temperature dependent in $\mu_0 = 1$\,T and essentially temperature-independent in $\mu_0H = 2$\,T. 
{\it This temperature-independence is inconsistent with a thermally activated relaxation process.}
The inset of Fig.\,\ref{relax}c shows an enlarged region of the plot. No correlation of relaxation with temperature is apparent.
The relaxation for $T >16$\,K is too fast to be measured directly and $M(t)$ is immediately constant once the field is stabilized. 
The relaxation process shows clear anomalies at the resonance fields (step-positions in $M$-$H$ loops) as shown in Supplementary Fig.\,\ref{SI-relax}.

{\bf Relaxation after switching off the field - Figs.\,\ref{relax}\,d-f.}
The relaxation is fast for $T = 16$\,K but suppressed for lower temperatures. However, the decrease is non-uniform with temperature and markedly smaller at the lowest temperatures. 
$M(t)$ can be fitted to a stretched exponential 
\begin{equation}
M(t) = M_0\,{\rm exp}[-(t/\tau)^\beta],
\end{equation}
with $M_0 = M(t=0)$ and $\tau$ is the relaxation time (see Supplementary Data\,\ref{si-fitting} for details). 
For $T > 16$\,K, $\tau$ is too small to be measured directly but can be determined from the out-of-phase part of the ac-magnetic susceptibility, $\chi''(T)$. 
The inset in Fig.\,\ref{relax}f shows maxima in $\chi''(T)$ at temperatures $T_{\rm max}$ which increase with the excitation frequency $f$.
The maximum in $\chi''(T)$ corresponds to a maximum in the energy absorption of the ac-field by the sample and shows that the relaxation time equals the time scale of the experiment, i.e., $\tau = 1/f$ at $T = T_{\rm max}$. 
The values obtained for $\tau$ are shown in Fig.\,\ref{relax}f in form of an Arrhenius-plot.
For $T \geq 16$\,K, $\tau$ follows a linear (Arrhenius-)behaviour with good agreement between directly and indirectly measured values. 
A fit of $\tau$ to a thermally activated law ($\tau = \tau_0\, {\rm exp}[\Delta E/k_{\rm B}T]$, dashed line) gives an energy barrier of $\Delta E /k_{\rm B} = 430$\,K and a pre-exponential factor $\tau_0 = 2.8\cdot10^{-10}$\,s. 
For $T < 16$\,K, $\tau$ deviates significantly from Arrhenius-behaviour and decreases much slower with decreasing temperature than expected for a thermally activated relaxation. 
A plateau, as shown in Fig.\,\ref{relax}f, has been found in SMMs\,\cite{Paulsen1995,Sessoli1997} and is a clear fingerprint of quantum tunnelling. 

{\bf Relaxation in negative fields - Figs.\,\ref{relax}\,g-i.}
This method is a variation of the previous one where the field is not simply switched off but ramped to a negative value opposite to the initial direction. 
As shown by the large magnetic hysteresis and the relaxation measurements above, the dominating energy scale of the system lies in the region of several Tesla acting on a few $\mu_{\rm B}$. 
Therefore, a magnetic field of a few milliTesla can, at most, merely change the energy level scheme.
Furthermore, applying a negative field is expected to accelerate the decrease of $M(t)$.
In contrast to both assumptions, small fields of $\mu_0H = -1$ to $-10$\,mT have a dramatic effect on the relaxation time and lead to an {\bf increase} of $\tau$ by several orders of magnitude (black curves in Fig.\,\ref{relax}h). 
In fact, the relaxation in $\mu_0H = -10$\,mT is becoming so slow that an accurate determination of $\tau$ is not possible within 4000\,s but requires much longer measurement times (which is beyond the scope of this publication). 
The main effect of the small negative field is likely a destruction of the tunnelling condition by lifting the zero-field-degeneracy - resonant tunnelling of the magnetization occurs only between degenerate states\,\cite{Sessoli2003}. 

In intermediate fields of $\mu_0H = -50$ to $-500$\,mT, the magnetization decreases rapidly until the applied field is stable (i.e. after 50\,s for $\mu_0H = -500$\,mT), followed by a significantly slower relaxation.
This indicates the presence of two time-scales for the relaxation, a fast one and a slow one.
The slow relaxation can be associated with overcoming the large energy barrier in the $a$-$b$ plane. However, the flipped moments are still not (anti-)parallel to the $c$-axis (Fig.\,\ref{model}). 
The fast relaxation can be associated with the subsequent, full alignment of the magnetic moments along the $c$-axis  by overcoming smaller, local maxima in the magnetic anisotropy energy (corresponding to the smaller steps in the $M$-$H$ loop, see Fig.\,\ref{model}).
Only larger fields of $\vert\mu_0H\vert > 1$\,T cause a faster decrease of $M(t)$ when compared with zero field (Fig.\,\ref{relax}i). 
A fast relaxation on the timescale of the experiment is reached in $\mu_0H = -7$\,T where $M(t)$ is saturated after $t \approx 2000$\,s. 
Accordingly, the magnetization reaches saturation in a similar short time after applying a field of $\mu_0H = 7$\,T to an unmagnetized sample (analogue to Figs.\,\ref{relax}a-c, not shown).
It should be noted that the time-dependencies, as presented in this section, are not restricted to low Fe concentrations. Clear relaxation effects are observed through the whole concentration range with the tendency to slow down with increasing $x$.

\section*{Discussion}
Our work shows that Li$_2$(Li$_{1-x}$Fe$_x$)N not only has a clear and remarkable anisotropy, generally not associated with Fe moments, but also shows time-dependence more consistent with SMM systems.
The strong correlation of relaxation time and coercivity field (Fig.\,\ref{relax}f) indicates that the slow relaxation leads to magnetic hysteresis for $x \ll 1$, i.e., the dilute system is not a ferromagnet and hysteresis emerges from a slow decay of a polarized, paramagnetic state. Whether this holds true for the dense system ($x = 0.28$), too, is not settled at this point - below $T \approx 65$\,K the ordering appears to be static on the timescale of M\"ossbauer spectroscopy\,\cite{Klatyk2002, Ksenofontov2003} (performed on polycrystalline material, $x = 0.16$ and 0.21).
  
\begin{figure}
\center
\includegraphics[width=0.9\textwidth]{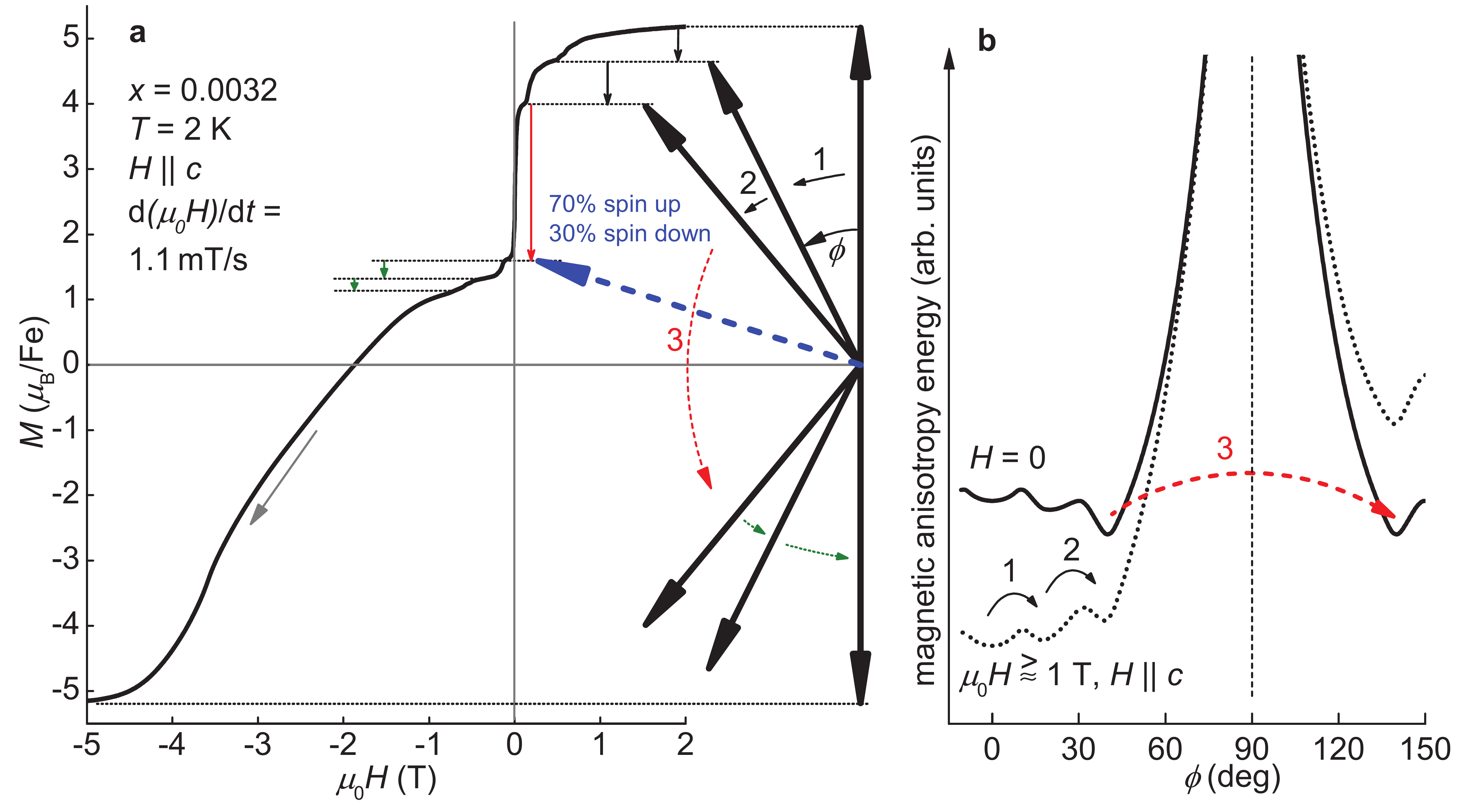}
\caption{Magnetic moment orientation and anisotropy energy. 
\textbf{a}, Magnetization of Li$_2$(Li$_{1-x}$Fe$_x$)N and a corresponding, possible orientation of the magnetic moment. This simplified model assumes a constant magnitude of the magnetic moment independent of the orientation with respect to the $c$-axis ($\phi$ is the angle between moment and $c$-axis).
The plateau after step 2 corresponds to $\phi\sim40^\circ$. The large jump in $M(H)$ at $H\sim0$ is in accordance with a reorientation of 30\,\% of the Fe magnetic moments from spin-up ($\phi\sim40^\circ$) to spin-down ($\phi\sim140^\circ$). 
A further, rapid reorientation for $H < 0$ is blocked because the resonance condition for magnetic tunnelling is destroyed for $H \neq 0$.
\textbf{b}, Possible schematic of the magnetic anisotropy energy as a function of the angle to the crystallographic $c$-axis. The global minimum in zero-field appears at an angle of $\phi\sim40^\circ$. Adding a sufficiently large Zeeman term for an applied field along the $c$-axis (dotted line) leads to a global minimum located at $\phi = 0$. 
The large step in the $M$-$H$ loop occurring at zero field is associated with the resonant tunnelling through the large energy barrier centred at $\phi = 90^\circ$.
Accordingly, the smaller steps correspond to transitions from a local to the (actual) global minimum in the total magnetic energy. 
\label{model}
}
\end{figure}

A possible schematic for the magnetic moment orientation and the magnetic anisotropy energy is depicted in Fig.\,\ref{model}. This model is purely based on the observed plateaus in the $M$-$H$ measurements and, beyond the extreme uniaxial symmetry of the iron environment, the origin of the magnetic anisotropy of Li$_2$(Li$_{1-x}$Fe$_x$)N is still unclear. 
The anisotropy found in SMMs is based on the magnetic interactions within the transition metal cluster or on the single ion anisotropy of lanthanide ions (or on both). 
For the latter case, the anisotropy is caused by the crystal electric field which acts as perturbation on the multiplet ground state of the lanthanide ion which is determined by Hunds rule coupling. 
Neither of these conceptionally simple scenarios is applicable for Li$_2$(Li$_{1-x}$Fe$_x$)N since the Fe-atoms are too dilute to strongly interact with each other for $x \ll 1$ and the crystal electric field is too strong to be regarded as a small perturbation when compared with the spin orbit coupling. 
Instead we are left with the complex determination of the energy level scheme of the relevant Fe-3$d$ electrons. 
Calculations based on LDA predict a counter-intuitive energy level scheme with the 3$d_z^2$ level having the lowest energy followed by partially occupied $d_{x^2-y^2}$, $d_{xy}$ levels\,\cite{Novak2002}. 
A very similar level scheme was recently found for a mononuclear Fe-based SMM\,\cite{Zadrozny2013p} sharing a similar structural motif as the Fe site in Li$_2$(Li$_{1-x}$Fe$_x$)N: a linear Fe$^{+1}$ complex, which seems to be an essential ingredient for the emergence of unquenched orbital moments and large magnetic anisotropy.

Even though the exact microscopic origin of the energy barriers separating the magnetic states in Li$_2$(Li$_{1-x}$Fe$_x$)N is not yet understood, we argue that quantum tunnelling has to be invoked to overcome them. 
From the Arrhenius-fit we find $\Delta/k_{\rm B} = 430$\,K (Fig.\,\ref{relax}f). 
Magnetization measurements on single crystals allow for a second method to estimate the barrier-height: 
when $H$ reaches the anisotropy field all moments are aligned along the field and the energy barrier is overcome. This gives rise to $\Delta E/k_{\rm B} \approx \mu_0H_{\rm ani} \cdot \mu_{\rm sat}/k_{\rm B} \approx 88\,{\rm T} \cdot 5\,\mu_{\rm B}/k_{\rm B} = 300$\,K in reasonable agreement with the above result. This value also agrees with the LDA based calculation of the anisotropy energy for $x = 0.17$ ($\Delta/k_{\rm B} = 278$\,K)\,\cite{Novak2002}.  
Such a large barrier cannot be overcome thermally at $T = 2$\,K - an estimation of the relaxation time from thermally activated law and $\Delta/k_{\rm B} = 278$\,K yields $\tau > 10^{43}$\,years (give or take a millennium). 
Therefore, a tunnelling process is likely involved in the relaxation process where the applied magnetic field shifts the energy levels in and out of degeneracy in full analogy to SMMs.
At this point it is not clear which levels of the single ion are brought in resonance by the applied magnetic field to explain the smaller steps in the $M$-$H$ loops.
Hyperfine interactions, as invoked for lanthanide based SMMs\,\cite{Ishikawa2007} and LiY$_{0.998}$Ho$_{0.002}$F$_4$\,\cite{Giraud2001}, are likely too weak to account for steps at applied fields of $\mu_0H = 0.5$\,T in particular since 98\% of the Fe-atoms do not carry a nuclear moment.

Energy barriers of similar ($\Delta/k_{\rm B} = 300$\,K\,\cite{Zadrozny2013p,Ishikawa2004}) or even higher size ($\Delta/k_{\rm B} = 800$\,K\,\cite{Gonidec2011}) have been observed in mononuclear SMMs. 
However, the remanent magnetization for these samples is very small caused by short relaxation times at low temperatures, i.e., $\tau$ deviates from Arrhenius-behaviour with a relaxation time at the plateau in the order of seconds - in contrast to $\tau \sim 10^5$\,s found for Li$_2$(Li$_{1-x}$Fe$_x$)N.
The fortunate combination of a deviation from Arrhenius behaviour at high temperatures ($T > 10$\,K) with large values at the plateau leads to the extreme coercivities presented here.
For potential data storage applications and stable magnetic materials, it is desirable to increase $\tau$ by suppressing the tunnel effect which can be achieved by enhanced exchange coupling\,\cite{Rinehart2011,Wernsdorfer2002}.  
On the other hand, studying tunnelling phenomena by itself requires isolated magnetic moments.
Both goals can be satisfied by varying the Fe concentration in Li$_2$(Li$_{1-x}$Fe$_x$)N accordingly. 
Thus for low Fe concentration of $x = 0.0032$, statistically, 98\,\% of the Fe atoms have only Li as nearest neighbours in the $a$-$b$ plane and Fe-Fe interactions are negligible. 
Larger Fe concentrations show higher coercivity fields at the lowest temperatures (Supplementary Fig.\,\ref{SI-mag}), which are caused by larger relaxation times, indicating that Fe-Fe interactions are indeed detrimental to tunnelling. Provided the energy barriers can be further enhanced and tunnelling appropriately controlled this opens a route for the  creation of hard permanent magnets from cheap and abundant elements. 

One remaining question is: why is quantum tunnelling so elusive in inorganic compounds? Besides our discovery in Li$_2$(Li$_{1-x}$Fe$_x$)N we are aware of only one other family of inorganic compounds showing macroscopic quantum tunnelling effects of the magnetization: LiY$_{0.998}$Ho$_{0.002}$F$_4$\,\cite{Giraud2003,Giraud2001} and related systems. However, the characteristic energy scales are two orders of magnitude smaller than in Li$_2$(Li$_{1-x}$Fe$_x$)N and magnetic hysteresis emerges only below $T = 200$\,mK with coercivity fields of $\mu_0H_{\rm c} = 30$\,mT.
To observe a macroscopic quantum effect such as tunnelling of the magnetization, the interaction between the magnetic moments has to be small. Very generally, coupling leads to excitations (modes), which lead to dissipation that destroys the quantum state. 
This rules out systems with dense, interacting moments.
Diluted systems of local magnetic moments have been the subject of extensive research mainly to study the Kondo effect. This necessarily requires metallic samples; the magnetic moments are not isolated but coupled to the electron bath which again leads to dissipation and the destruction of the quantum state. 
Insulating samples with diluted or non-interacting magnetic moments have been far less studied - diluted magnetic semiconductors are explicitly excluded from this statement because of their finite carrier density.
A possible explanation for the absence of magnetic tunnelling in insulators is a Jahn-Teller distortion which is frequently observed, e.g., in lanthanide zircons of the form $RX$O$_4$ (Ref.\,\cite{Kirschbaum1999} and references therein).
According to the Jahn-Teller theorem\,\cite{Jahn1937} the orbital degeneracy of Fe should cause a structural distortion also for Li$_2$(Li$_{1-x}$Fe$_x$)N in order to reach a stable state. In the diluted case it will occur locally for Fe and not necessarily for the whole crystal. 
Lifting the orbital degeneracy corresponds to a zero-orbital angular momentum, which is the only fully non-degenerate state. Consequently, a Jahn-Teller distortion would lead to a loss of the magnetic anisotropy and a decay of the energy-barrier.
However, the Fe atom sits between two nitrogen neighbours which provide the dominant bonding. These three atoms can be regarded as acting like a linear molecule which is not subject to a Jahn-Teller distortion\,\cite{Jahn1937}.

\noindent
To summarize, we demonstrated a huge magnetic anisotropy and coercivity in Li$_2$(Li$_{1-x}$Fe$_x$)N and want to emphasize the three properties which are probably essential for the emergence of macroscopic quantum tunnelling: (i) the compound is insulating, (ii) the orbital magnetic moment of Fe is not quenched, and (iii) the N-Fe-N complex forms a linear molecule avoiding a Jahn-Teller distortion.
These properties may serve as a basis for the design of materials featuring even higher characteristic energy scales.

\section*{Methods}

\subsection*{Crystal growth}
Starting materials were Li granules (Alfa Aesar, 99\%), Li$_3$N powder (Alfa Aesar, 99.4\,\%), and Fe granules (99.98\,\%). 
The mixtures had a molar ratio of Li\,:\,Fe\,:\,N = $9-x_0$\,:\,$x_0$\,:\,1 with $x_0$ = 0 to 0.5. A total mass of roughly 1.5\,g was packed into a 3-cap Ta-crucible\,\cite{Canfield2001} inside an Ar-filled glovebox. 
The Ta-crucible was sealed by arc-melting under inert atmosphere of $\sim$\,0.6\,bar Ar and subsequently sealed in a silica ampoule. 
The Li-Fe-N mixture was heated from room temperature to $T = 900^\circ$C over 4\,h, cooled to $T = 750^\circ$C within 1.5\,h, slowly cooled to $T = 500^\circ$C over 62\,h and finally decanted to separate the Li$_2$(Li$_{1-x}$Fe$_x$)N crystals from the excess liquid.
Single crystals of hexagonal and plate-like habit with masses $>$100\,mg could be obtained. The maximum lateral sizes of $\approx$10\,mm were limited by the crucible size where the crystal thickness is typically $\approx$1\,mm.
Typically we found a few large single crystals with similar orientation clamped between the container walls above the bottom of the crucible and also several smaller ones attached to the bottom. 
We used pieces of the larger crystals for the magnetization measurements presented in this publication.
\\
It should be noted that the smallest Fe concentration of $x = 0.00028$, as measured by inductively coupled plasma mass spectrometry (ICP-MS, see below), was obtained without intentional introduction of Fe in the melt. Fe was most likely introduced as impurity from the starting materials or from the Ta-crucible.

\subsection*{Chemical analysis - ICP-MS}
The Li$_2$(Li$_{1-x}$Fe$_x$)N samples were analysed using an inductively coupled plasma magnetic sector mass spectrometer (ICP-MS, Element 1, Thermo Scientific). The samples were introduced into the ICP via a low-flow nebulizer (PFA-100, Elemental Scientific Inc.) and double-pass spray chamber. The interface between the ICP and mass spectrometer was equipped with nickel sampler and skimmer (H-configuration) cones.  
The mass spectrometer was operated in medium resolution ($m/\Delta m = 4000$) to separate the Fe$^+$ isotopes of interest from interfering species.  
The detector was operated in dual mode allowing for the operating software to selectively switch between analogue and counting measurements. 
Prior to sample analysis, the torch position and instrumental operating parameters (Table.\,\ref{icpms-tab}) were adjusted for maximum peak height and signal stability.
The main elements of interest for quantification were lithium (mass to charge ratio $m/z$\,7) and iron ($m/z$\,55.935 and 56.935). Tantalum ($m/z$\,181) and calcium ($m/z$\,43 and 44) were also measured to check for contamination from the crucible material and known impurities of the starting materials.  
Following these initial analyses, a full isotopic spectrum ($m/z$\,7 to $m/z$\,238) was measured in low resolution ($m/\Delta m = 300$) for several representative samples to check for any other possible sources of contamination.
Carbon, nitrogen, and oxygen could not be measured due to high background levels contributed by the acid solution, the argon gas, and the instrumental components.
Minor amounts of tantalum and calcium were measured in the Li-Fe-N samples. All samples contained less than 0.4\,mass\,\% calcium (corresponding to Li$_{1-\delta}$Ca$_\delta$ with $\delta < 0.007$) and less than 0.1\,mass\,\% tantalum (corresponding to Li$_{1-\delta}$Ta$_\delta$ with $\delta < 0.0004$). The combined concentrations of all other contaminant elements comprised far less than 0.01\,mass\,\% of the solid samples.

\begin{table}
\center
\caption{ICP-MS operating parameters}
\begin{tabular}{ll}
\hline
\hline
RF power & 1150\,W\\
\hline
Outer gas flow & 16\,l\,min$^{-1}$\\
\hline
Auxiliary gas flow & 1.5\,l\,min$^{-1}$\\
\hline
Sample gas flow (argon) & 1.04\,l\,min$^{-1}$\\
\hline  
Sampling position & 13\,mm from load coil, on center\\
\hline
Signal ratios for tuning and optimization & Ce$^{2+}$/Ce$^+ \approx 4$\,\%, CeO$^+$/Ce$^+ \approx 7$\,\%\\
\hline
\hline    
\end{tabular}
\label{icpms-tab}
\end{table}

All of the Li$_2$(Li$_{1-x}$Fe$_x$)N samples were dissolved for ICP-MS analysis. Approximately 5 to 25\,mg of solid sample was weighed accurately into an acid vapour washed Teflon bottle on a balance. 
A small amount (2 to 4\,g) of cold ($\sim 3^\circ$C) deionized water was added and allowed to react. 
Once the sample mass stabilized, approximately 1.5\,g of 70\,\% nitric acid was added to completely dissolve the remaining solid.  Upon complete dissolution, the solution was diluted with deionized water to a mass of 50\,g. 
Aliquots of these original solutions were diluted with prepared aqueous 1\,\% nitric acid to a concentration of 1 to 5\,ppm in terms of the original solid sample mass.
Standard solutions were prepared for the quantification of iron and lithium in the samples. A 1\,ppm iron and lithium standard was prepared by diluting 1000\,ppm stock solutions (SPEX CertiPrep, High-Purity Standards) with cleaned 1\,\% nitric acid.  
Lower concentration standards were prepared via dilutions of the original 1\,ppm standard solution. Blanks of the water and acids were analysed and had negligible amounts of the analyte elements. 
The water used was 18\,M$\Omega$\,cm (Barnstead Nanopure) and the nitric acid was purified by sub-boiling distillation (Classic Sub-boiling Still Assembly, Savillex) before use.   
Lithium is prone to memory effects in ICP-MS due to either sample introduction or deposition and vaporization of Li from the cones. The Li$^+$ (and Fe$^+$) signals from the samples and standards all rinsed out to baseline between the measurements.

% \bibliography{C:/Users/jesche/Documents/Prasentationen/paper/zitate}

\begin{appendix}
\noindent{\bf Acknowledgements} 
Bruce Harmon, Yongbin Lee, Natalia Perkins, Yuriy Sizyuk, Vladimir Antropov, Makariy Tanatar, Hyunsoo Kim, Ruslan Prozorov, and Yuji Furukawa are acknowledged for comments and discussions. The authors thank Gregory Tucker for assistance with recording Laue-back-reflection pattern, Jakoah Brgoch for assistance with early X-ray powder diffraction measurements, and Jim Anderegg for discussions and first attempts of performing Auger-spectroscopy on these samples. Kevin Dennis is acknowledged for assistance with magnetization measurements. 
This work was supported by the U.S. Department of Energy, Office of Basic Energy Science, Division of Materials Sciences and Engineering. The research was performed at the Ames Laboratory. Ames Laboratory is operated for the U.S. Department of Energy by Iowa State University under Contract No. DE-AC02-07CH11358.
\\
\noindent{\bf Supplementary Information} is attached below.
\\
\noindent{\bf Author Contributions} A.\,J and P.\,C developed the Li-N growth technique and initiated this study. 
A.\,J. grew the single crystals and performed the magnetization measurements. 
S.\,T. collected and analysed single crystal X-ray diffraction data.
J.\,J. and R.\,H. performed the chemical analysis. 
P.\,C, B.\,M., S.\,B., V.\,T., and A.\,J. analysed and interpreted the magnetization data.
A.\,K. and A.\,J. collected and analysed powder X-ray and Laue-back-reflection data. 
A.\,J. and P.\,C. wrote the manuscript with the help of all authors. 
\\
\noindent{\bf Competing Interests} The authors declare that they have no competing financial interests.
\\
\noindent{\bf Correspondence} should be addressed to A.\,Jesche (email:jesche@ameslab.gov).
\end{appendix}

\clearpage

\setcounter{figure}{0}
\setcounter{table}{0}
\renewcommand{\figurename}{Supplementary Figure}
\renewcommand{\tablename}{Supplementary Table}

\section*{Supplementary Figures}
\vspace{+8ex}

\begin{figure}[!ht]
\center
\includegraphics[width=0.666\textwidth]{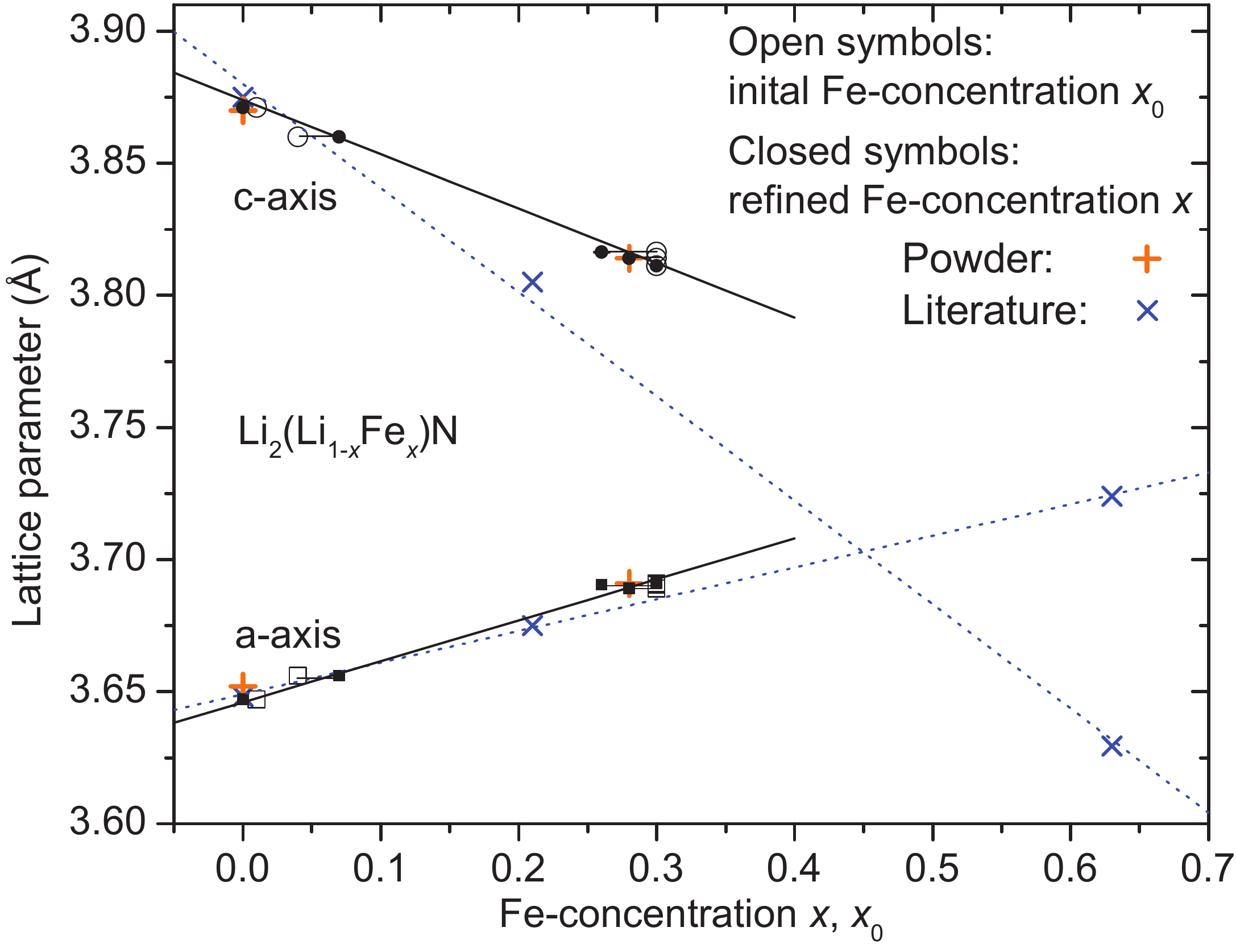}
\caption{Structural parameters determined by single crystal X-ray diffraction.
The single crystal data is plotted twice, once as a function of the nominal \textit{x}$_0$ value and a second time using the refined \textit{x} value (see main article Fig.\,\ref{crystal}d).
The two coupled data points are linked with a small vertical bar. 
The refined values are in excellent agreement with the ones obtained by means of powder X-ray diffraction on ground single crystals (\textit{x}  = 0, and \textit{x} = 0.28, the Fe concentration of the powder was measured by inductively coupled plasma mass spectroscopy).  
The variation of lattice parameters for Li$_2$(Li$_{1-x}$Fe$_x$)N with respect to the refined Fe concentration follow Vegards law. 
The lattice parameters from the literature for \textit{x} = 0\,\cite{Rabenau1976} \textit{x} = 0.21\,\cite{Klatyk2002}, \textit{x} = 0.63\,\cite{Klatyk1999} are shown for comparison. Furthermore, there is excellent agreement with the lattice parameters given in Ref.\,\cite{Yamada2011} (not included for clarity).
\label{sc}
}
\end{figure}

\begin{figure}
\center
\includegraphics[width=0.6\textwidth]{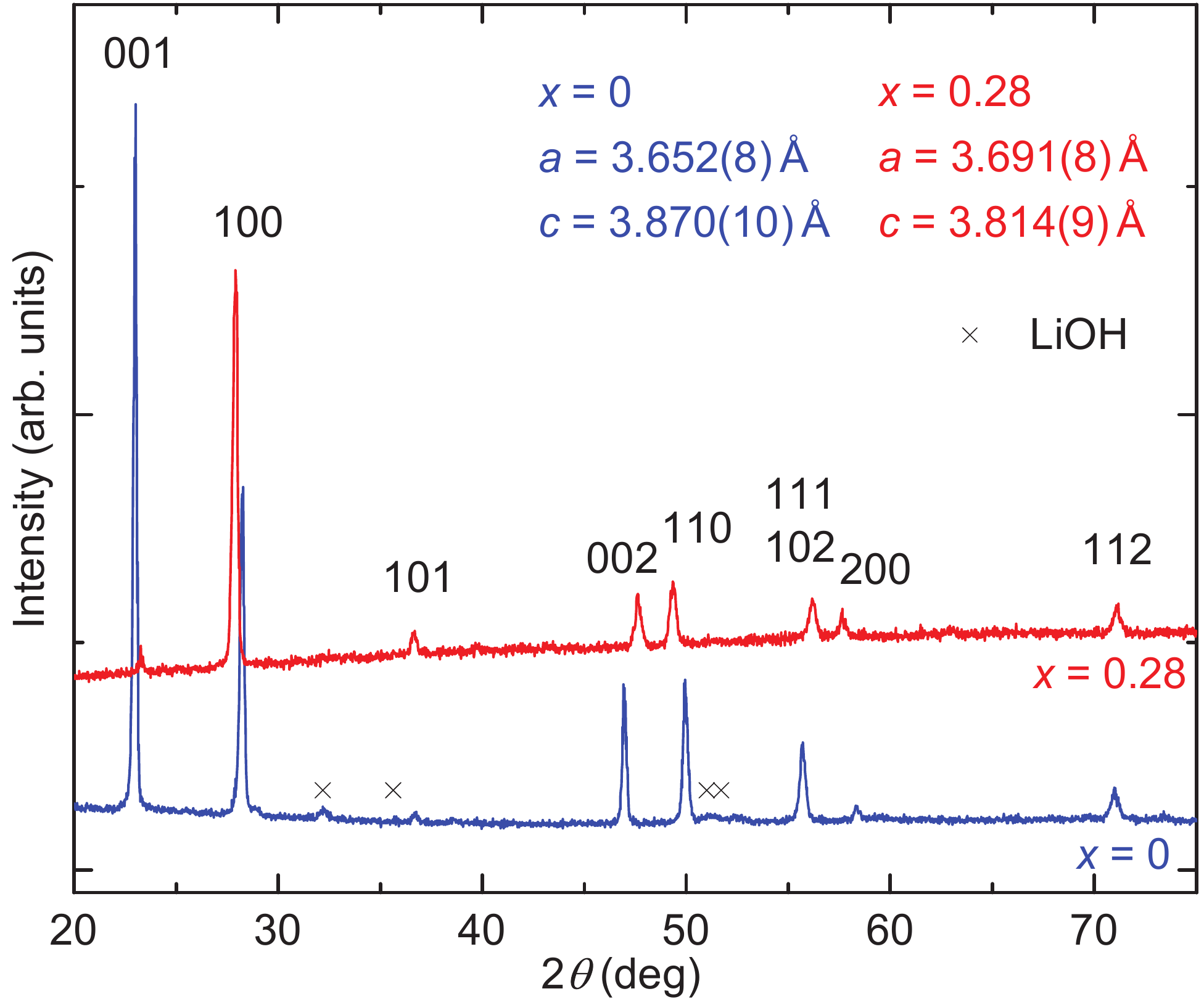}
\caption{X-ray powder diffraction pattern of Li$_2$(Li$_{1-x}$Fe$_{x}$)N for \textit{x} = 0 and \textit{x} = 0.28 recorded in a nitrogen filled glovebox. 
Each pattern was recorded within 2\,h using similar amounts of powder. The significantly larger background of the Fe-rich sample is caused by the fluorescence radiation of Fe.
The additional small intensity peaks in \textit{x} = 0 can be indexed based on the crystal structure of LiOH\,\cite{Mair1978} and emerge also for \textit{x} = 0.28 for extended measurement times. The intensity of these peaks increases with ongoing measurement time and is larger for fine ground powder.
\label{powderdiff}
}
\end{figure}

\begin{figure}
\center
\includegraphics[width=\textwidth]{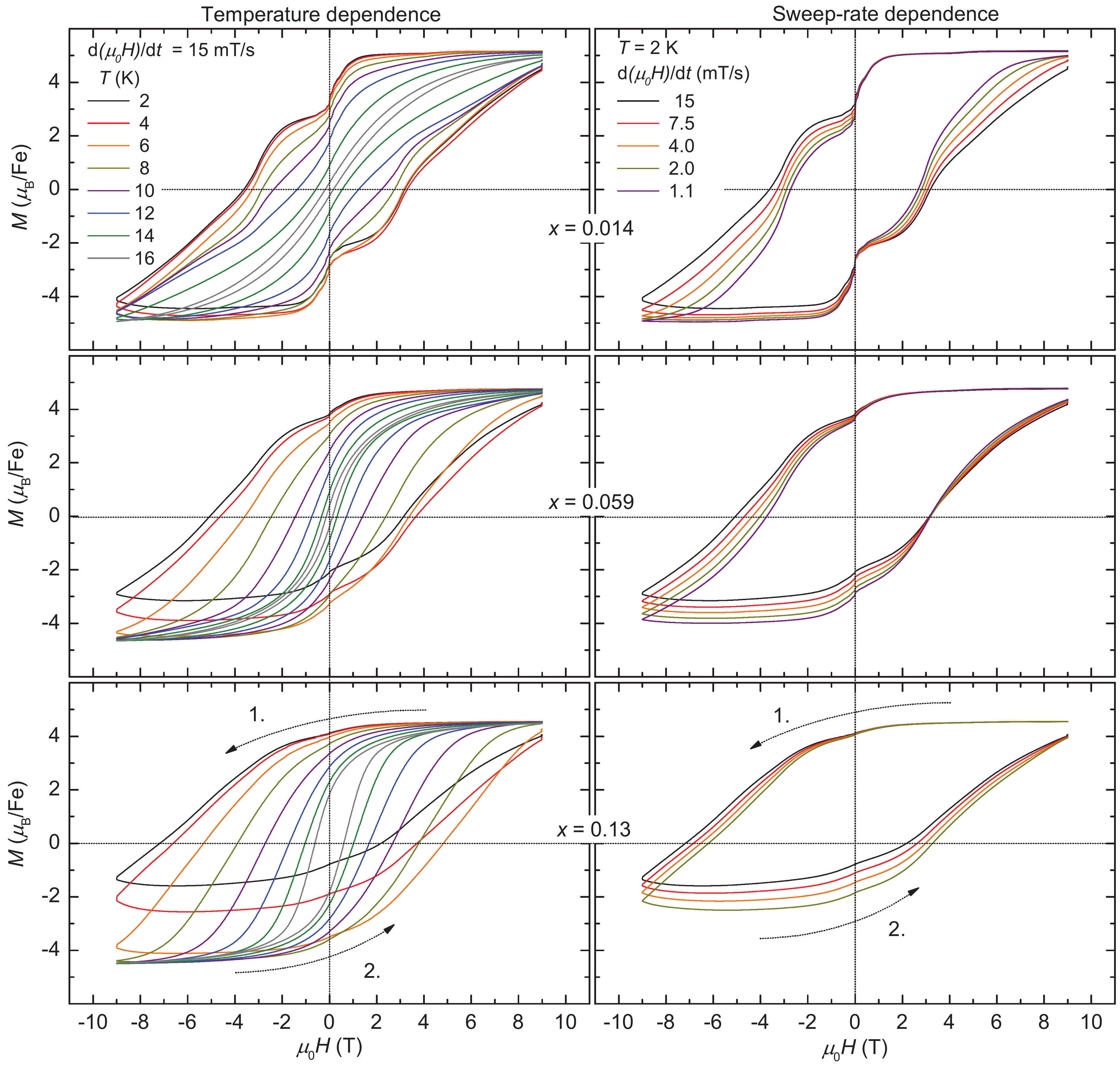}
\caption{Magnetization of Li$_2$(Li$_{1-x}$Fe$_x$)N for different \textit{x} for \textit{H}\,$\parallel$\,\textit{c}. The hysteresis loops depend on the temperature (left panels) as well as on the sweep-rate of the applied magnetic field (right panels) - see Supplementary Note\,4 for details.
The step at \textit{H} = 0 is subsequently suppressed with increasing \textit{x} indicating a detrimental effect of the enhanced magnetic coupling between the Fe magnetic moments to the tunnel effect.
Whether \textit{M} goes to zero or asymptotically approaches a finite value for \textit{t} $\rightarrow \infty$ in \textit{H} = 0 is not settled at this point. 
However, the decreasing size of the steps together with the increasing coercivity fields indicate the formation of a true ferromagnetic state for large enough Fe concentrations in contrast to a polarized paramagnetic state for \textit{x}~$\ll$~1.
\label{SI-mag}
}
\end{figure}

\begin{figure}
\center
\includegraphics[width=\textwidth]{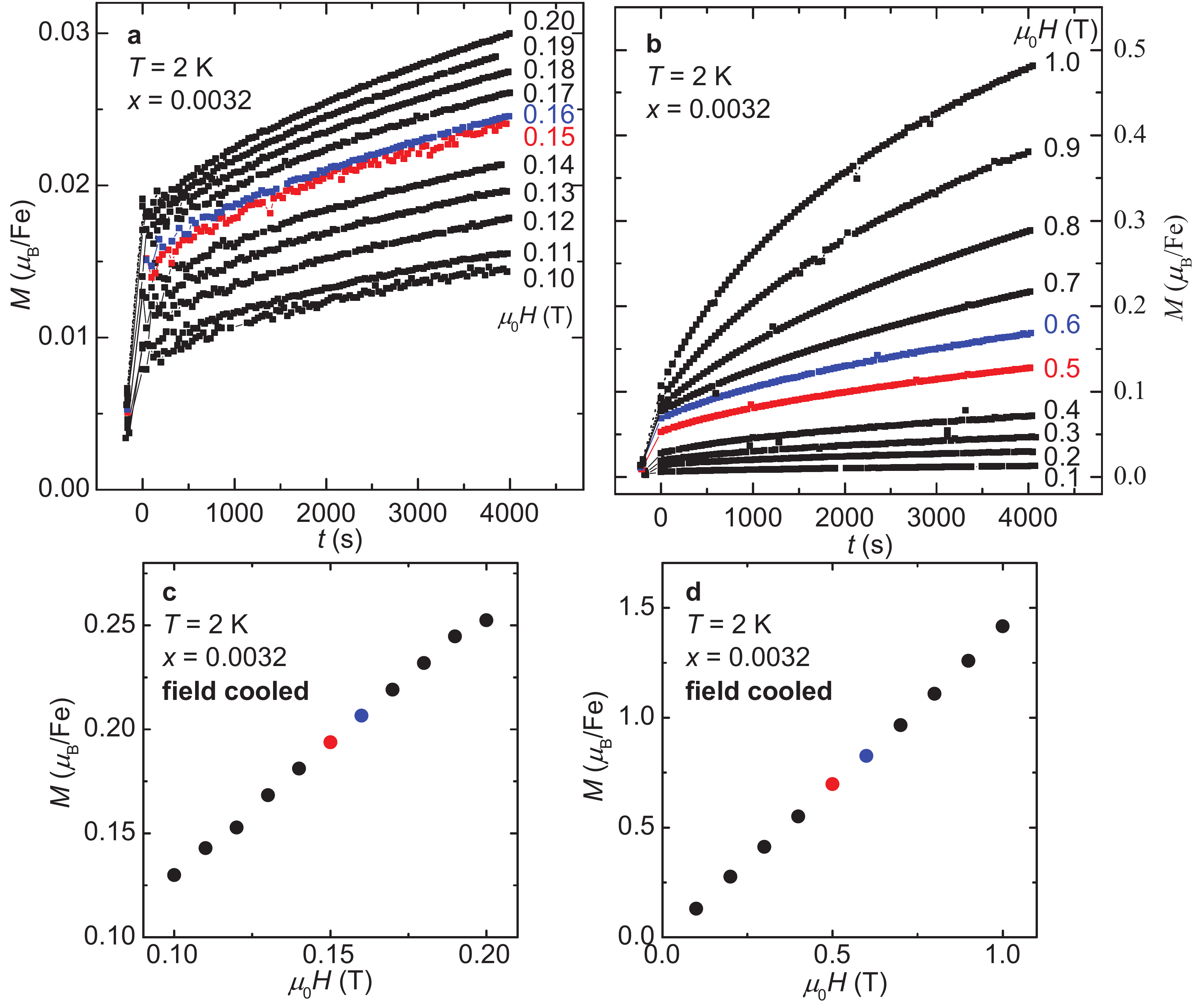}
\caption{Correlation of relaxation time and step-position in \textit{M}-\textit{H} loops of Li$_2$(Li$_{1-x}$Fe$_x$)N. 
The increase of the magnetization as a function of time is plotted for applied fields around the step-positions in the \textit{M}-\textit{H} loops. Stable fields of the values given in the plots are reached at \textit{t} = 0.
\textbf{a,} The relaxation depends on \textit{H} in a very non-linear fashion. For \textit{H}-values close to the step at $\mu_0$\textit{H} = 0.15\,T the relaxation becomes field-independent.
\textbf{b,} For \textit{H}-values close to the smaller step at $\mu_0$\textit{H} = 0.55\,T the non-linearity of the field-dependence is less pronounced.
However, the increase in \textit{M(t)} for increasing the applied field from $\mu_0$\textit{H} = 0.4\,T (off-resonant) to 0.5\,T (on-resonant) is still significantly larger than the difference between applied fields of $\mu_0$\textit{H} = 0.3\,T and 0.4\,T or $\mu_0$\textit{H} = 0.6\,T and 0.7\,T.
{\bf c,d} The corresponding field-cooled measurements show a linear dependence of the magnetization on the applied field indicating that the non-linear behavior observed in the relaxation is indeed a dynamic property (see Supplementary Note\,4 for details).
\label{SI-relax}
}
\end{figure}

\begin{figure}
\center
\includegraphics[width=0.9\textwidth]{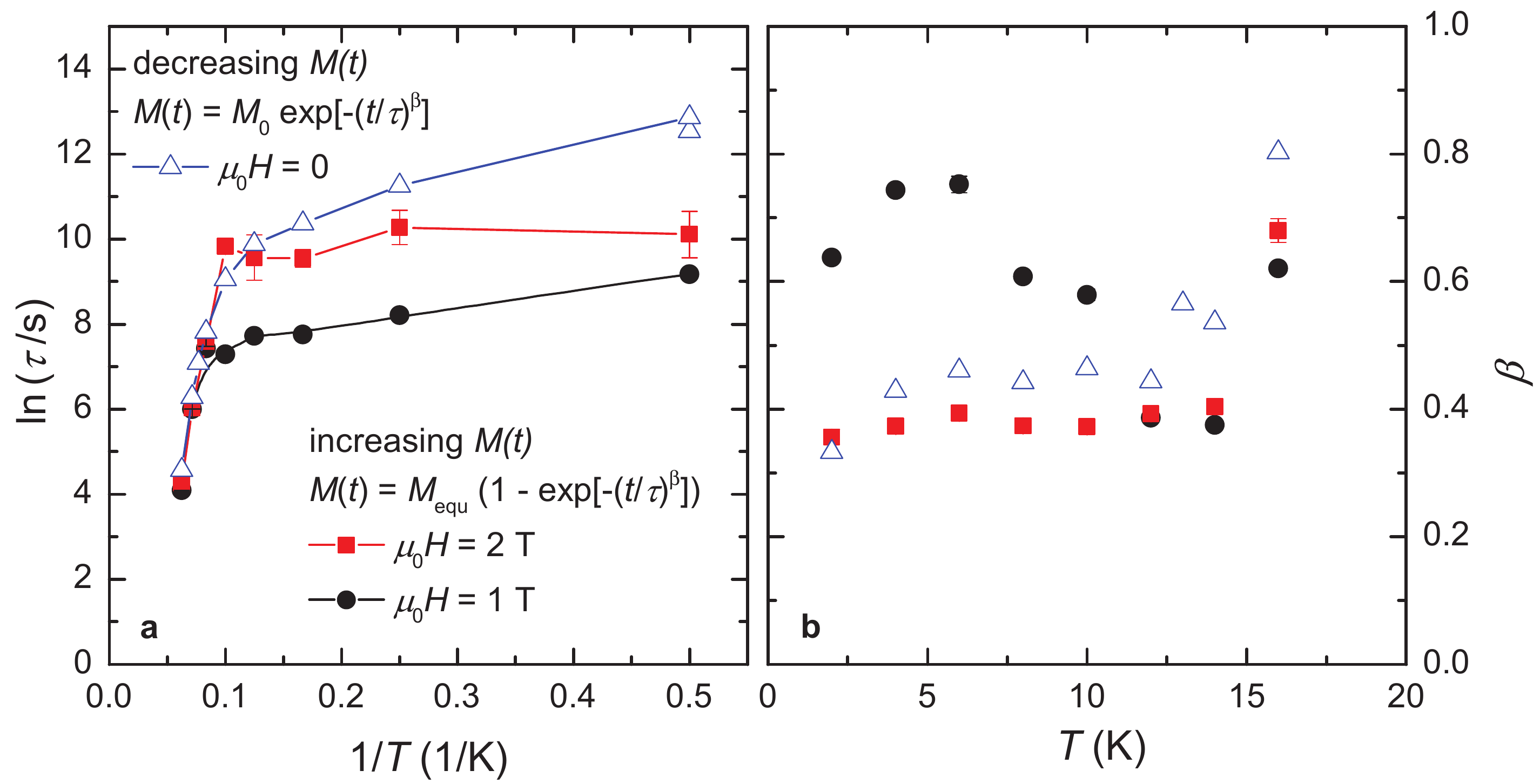}
\caption{Time-dependent magnetization analysed by stretched exponential functions. \textbf{a,} A roughly temperature-independent relaxation time was found in an applied field of $\mu_0H = 2$\,T for $T < 10$\,K (see Supplementary Note\,4 for details). The relaxation in $\mu_0H = 1$\,T and the decay of the magnetization in zero-field show similar behaviour where the relaxation times increase only weakly with decreasing temperature. (The standard errors of the fit are indicated or appear smaller than the symbol size.)  
\textbf{b,} The exponent $\beta$ accounts for a change of the relaxation rate during time with a faster relaxation at early times. For the relaxation in zero field, $\beta$ tends to increase with increasing temperature and approaches $\beta = 1$ (corresponding to one single relaxation time). No clear trend is observed for relaxation in applied fields.
\label{SI-tau}
}
\end{figure}

\clearpage

\section*{Supplementary Tables}

\begin{table}[h]
\center
% \captionsetup{width=.7\textwidth}
\caption{Crystal data and structure refinement for Li$_2$(Li$_{0.74}$Fe$_{0.26}$)N}
\begin{tabular}{ll}
\hline
\hline
Empirical formula & Li$_2$(Li$_{0.74}$Fe$_{0.26}$)N\\ 
\hline
Formula weight & 47.55\,g\,mol$^{-1}$\\
\hline
Wavelength & 0.71073\,\AA\\
\hline
Crystal system, space group & Hexagonal, \textit{P}\,6/\textit{m\,m\,m} (\#191)\\
\hline
Unit cell dimensions & \textit{a} = 3.6904(2)\,\AA \\
  & \textit{c} = 3.8164(2)\,\AA \\   
\hline
Volume & 45.012(4)\,\AA$^3$ \\
\hline
Z, Calculated density & 1, 1.754\,g\,cm$^{-3}$ \\
\hline
Absorption coefficient & 2.073\,mm$^{-1}$ \\
\hline
Crystal size & 0.11 $\times$ 0.08 $\times$ 0.05\,mm$^3$ \\
\hline
$\theta$\,range & 5.34$^\circ$ to 52.02$^\circ$ \\
\hline
Limiting indices & -8 $\leq$\textit{h} $\leq$ 5 \\
~ & -5 $\leq$ \textit{k} $\leq$ 7 \\
~ & -8 $\leq$ \textit{l} $\leq$ 6 \\
\hline
Reflections collected/unique & 1176/138 [\textit{R}(int) = 0.0172] \\
\hline
Absorption correction & Empirical \\
\hline
Data/parameters & 138/8 \\
\hline
Goodness-of-fit on $\vert$\textit{F}$\vert ^2$ & 1.154 \\
\hline
Final $R$ indices [\textit{I} $ > 2\sigma$(\textit{I})] & \textit{R}1 = 0.0323, w\textit{R}2 = 0.0927 \\
\hline
\hline
\end{tabular}
\label{sc-tab1}
\end{table}

\begin{table}[b]
\center
% \captionsetup{width=.73\textwidth}
\caption{Atomic coordinates and isotropic displacement parameters for Li$_2$(Li$_{0.74}$Fe$_{0.26}$)N. $U_{\rm eq}$ is defined as one third of the trace of the orthogonalized $U_{\rm ij}$ tensor. 
}
\begin{tabular}{ccccccc}
\hline
\hline
Atom & Wyckoff site & Occu. & ~~\textit{x}~~ & ~~\textit{y}~~ & ~~\textit{z}~~ & ~~~~\textit{U}$_{\rm eq}$(\AA$^2$)~~ \\
\hline
\hline
Fe1 & 1\textit{b} & 0.26(1)  & 0 & 0 & 1/2 & 0.010(1) \\
\hline
Li1 & 1\textit{b} & 0.74(1)  & 0 & 0 & 1/2 & 0.010(1) \\
\hline
Li2 & 2\textit{c} & 1        & 1/3&2/3&0 & 0.031(1) \\
\hline
N1 & 1\textit{a} & 1 & 0 & 0 & 0 & 0.010(1) \\
\hline
\hline
\end{tabular}
\label{sc-tab2}
\end{table}

\clearpage
\setcounter{section}{0}
\section*{Supplementary Note 1: Arguments against clustering of Fe atoms}\label{phase}
We want to point out that phase separation (partial segregation) or small cluster formation can not be definitely excluded in these materials, but the preponderance of data support the idea that single Fe atoms (and not clusters) act as mononuclear magnetic centres. Here we summarize some of the key indications pointing against Fe clustering or phase separation.
\\
Argument 1 - based on magnetic properties. The effective moment is essentially independent of the Fe concentration (main text Fig.\,\ref{mag-intro}d). This is incompatible with the formation of clusters consisting of Fe on adjacent Li-1$b$ sites, e.g., Fe-trimers or any particular configuration built from several neighboring Fe atoms in the $a$-$b$ plane. 
These would be diluted and non-interacting for $x \ll 1$ but so dense for $x \sim 0.3$ that a substantial overlap of randomly arranged clusters would occur. The peculiar magnetic properties of an isolated trimer (e.g. superparamagnetism) should be significantly different from a dense, interacting system. Accordingly, a change of the effective moment with the Fe concentration would be expected. This is not observed. However, covering the $a$-$b$ plane with non-touching Fe dimers or with Fe-N-Fe chains along the crystallographic $c$-axis is possible even for the largest investigated Fe concentration of $x = 0.28$. If, as reported in Ref.\,\cite{Klatyk1999}, substantial higher Fe concentrations can be grown this line of argument could be tested and analysed more rigorously.
\\
Argument 2 - based on magnetic properties. Samples with low Fe concentrations, $x = 0.0032$ and $x = 0.014$ (main article Fig.\,\ref{M-H}a and Supplementary Fig.\,\ref{SI-mag}, respectively) as well as $x = 0.00028$ (not shown), show similar $M$-$H$ loops. 
If the magnetic centres are allowed to be of variable size, e.g. monomer, dimer and trimer, or just monomer and dimer, then it would be reasonable to assume that the distribution of the cluster size should change significantly with $x$. Accordingly the magnetic properties, such as step-size and step-position in the $M$-$H$ loops, would be expected to strongly change. This is not observed. From this we conclude a monodisperse distribution of the magnetic centres.  
\\
Argument 3 - based on X-ray diffraction. Single crystal diffraction on the $x = 0.07$ sample did not show any indications for phase separation (segregation). A driving force for phase separation should be stronger in this concentration range when compared with smaller $x$. In other words: if a single phase, mixed crystal forms for $x =0.07$ than it should form for lower Fe concentrations as well.  
\\
Argument 4 - based on kinetic aspects of the crystal growth. When we diluted the Fe concentration down to $x \sim 0.005$ a cluster formation seems unlikely. If regions with a fixed Fe concentration (e.g. $x = 1/3$ or $x \sim 0.3$) would grow preferentially then the Fe should be concentrated in a small region of the growth whereas the majority of the product is expected to contain no Fe at all. 
More specifically, an initial melt with the composition Li:Fe:N = 8.995:0.005:1 and total mass of 1.3\,g is sufficient for the formation $\sim$\,7\,mm$^3$ or $\sim$\,13\,mg of '$x = 1/3$-material'. 
However, the actual yield of such growths was found to be $>$200\,mm$^3$. ICP-MS as well as magnetization measurements performed on multiple samples with low Fe concentration with $m \sim 10$\,mg, each, show Fe concentrations in the order of the initial Fe concentration $x_0$ and never revealed Fe concentrations of $x = 1/3$ or $x \sim 0.3$.
Therefore, there seems to be no driving force for a clustering or phase separation on a meso- or macroscopic scale.
\\
Argument 5 - based on M\"ossbauer spectroscopy.
The distribution of Fe-sites (the number of Fe atoms as nearest neighbor) as obtained by M\"ossbauer spectroscopy indicates a statistical substitution of Li by Fe for $x = 0.16$ and $x = 0.21$\,\cite{Klatyk2002,Ksenofontov2003}. It seems reasonable to assume a statistical distribution also for lower Fe concentrations. 
\\
To summarize, arguments 1 and 5 indicate the absence of randomly arranged, small clusters with Fe on adjacent Li-1$b$ sites (e.g. trimers or hexamers), argument 2 indicates a monodisperse collection of magnetic centres, and arguments 3 and 4 indicate the absence of phase separation on meso- and macroscopic length scales. 
Single Fe atoms as mononuclear magnetic units seem a valid operating assumption and the simplest and consistent description of our observations.

\section*{Supplementary Note 2: X-ray powder diffraction}\label{si-note1}
In accordance with the insulating behaviour and the proposed predominantly ionic bonding the samples are brittle and grind well into a fine powder.
The powder is dark brown in colour and turns white immediately after exposure to air. X-ray powder diffraction revealed LiOH which further decays into LiOH-H$_2$O as decay products in agreement with the literature data\,\cite{Gregory2001}. 
The diffraction patterns shown in Supplementary Fig.\,\ref{powderdiff} were recorded in a nitrogen filled glovebox using a Rigaku Miniflex diffractometer (wavelength: Cu-$K\alpha_{1,2}$). 
The Fe concentration was determined by chemical analysis of a larger crystal ($m \sim 25$\,mg), from which a smaller piece ($m \sim 10$\,mg) had been ground to the powder used for X-ray diffraction. 
The ground powders are very reactive and the decay rate of the powder, even in a nominally high purity nitrogen atmosphere, strongly depends on the grain-size (surface area). 
Grinding the material to a fine powder causes a faster decay, whereas a course powder results in unrealistic peak-profile and intensities.
In this sense, the diffraction pattern in Supplementary Fig.\,\ref{powderdiff} present an optimum between foreign phase formation and peak-profile. 
For this reason, we refrain from refining the powder-data, however, it is noteworthy that the strong suppression of the 0\,0\,1 reflection for $x = 0.28$ agrees well with the theoretically expected intensity of this peak related to the partial substitution of Li by Fe.
Furthermore, the significantly larger background of the Fe-rich sample reflects the fluorescence radiation from Fe.  

\section*{Supplementary Note 3: X-ray single crystal diffraction}\label{si-note2}

Laue back-reflection patterns on large single crystals (main article Fig.\,\ref{crystal}\,b) were taken with an MWL-110 camera manufactured by Multiwire Laboratories. An estimate to the X-ray florescent background was removed from the measured Laue pattern by fitting, and subsequently subtracting, a two-dimensional Gaussian. 
\\
Several smaller single crystals (dimensions $< 0.15$\,mm) were selected from crushed single crystals and sealed in thin-walled glass capillaries in a glovebox under nitrogen atmosphere ($<0.5$\,ppm O$_2$, H$_2$O). 
Diffraction intensities were collected at room temperature on a SMART APEX II diffractometer equipped with a CCD area detector using graphite monochromated Mo-$K\alpha$ radiation. 
The crystal-to-detector distance was 4.0\,cm. 
The diffraction data collection strategies were obtained from an algorithm in the program COSMO in the APEX II software package\,\cite{Bruker2013} composed of $\omega$ and $\phi$ scans. 
The step-size was $0.5^\circ$ and the exposure time was 10 to 30\,s/frame. Data were indexed, refined, and integrated with the program SAINT in the APEX II package\,\cite{Bruker2013}. 
An empirical absorption correction was performed using SADABS as implemented in the APEX II package\,\cite{Bruker2013}. 
The structure was solved via direct methods using the SHELXS program\,\cite{Sheldrick1997} and refined by full-matrix least squares on $\vert F \vert ^2$ for all data using SHELXL\,\cite{Sheldrick1997}. 
The variation of lattice parameters for Li$_2$(Li$_{1-x}$Fe$_x$)N with respect to Fe concentration (initial and refined composition) is shown in Supplementary Fig.\,\ref{sc}.  
Over all, the refined Fe concentrations are in reasonable agreement with the initial concentration $x_0$ and agree very well with the composition obtained from the ICP-MS analysis.
(The larger refined value of $x = 0.07$ for $x_0 = 0.04$ is in agreement with the trend observed for $x_0 \lesssim 0.1$, see main article Fig.\,\ref{crystal}d.) 
Furthermore, there is a good agreement with the lattice parameters obtained from powder X-ray diffraction and with the literature data shown in Supplementary Fig.\,\ref{sc} as well as with Ref.\,\cite{Yamada2011} (not included in the figure for clarity). 
Representative crystallographic data and details on the structural refinement are tabulated in Tables\,\ref{sc-tab1},\,\ref{sc-tab2} for one of the specimen (refined concentration $x = 0.26$, initial concentration $x_0 = 0.30$).

\section*{Supplementary Note 4: Magnetization measurements}\label{SI-mag2}
Magnetization measurements were performed using a Quantum Design 7T SQUID Magnetic Property Measurement System (main article Figs.\,\ref{mag-intro}a,c, Figs.\,\ref{relax}b,c, and Supplementary Fig.\,\ref{SI-relax}),
 a Quantum Design 9T Physical Property Measurement System equipped with a Vibrating Sample Magnetometer (main article Fig.\,\ref{M-H}, Figs.\,\ref{relax}e,\,h,\,i, Fig.\,\ref{model} and Supplementary Fig.\,\ref{SI-mag}), 
and a Quantum Design 14T Physical Property Measurement System equipped with an ACMS option (main article Fig.\,\ref{mag-intro}b using dc extraction and inset in Fig.\,\ref{relax}f using ac susceptibility).
Test measurements revealed identical magnetization values within the error range of $\sim 3$\,\% using the different setups on the same sample. 

The errors of effective moment and saturation moment for $H \parallel c$ given in main article Fig.\,\ref{mag-intro}d are estimated based on the errors in assessing the sample mass and the Fe concentration. For low Fe concentration, $x = 0.0032$, the diamagnetic contribution of the Li$_3$N host significantly effects the magnetization for $T > 100$\,K. 
The maximum host contribution is estimated from the diamagnetic core increments (Li$^{1+} \approx -0.2\cdot 10^{-10}$\,m$^3$mol$^{-1}$\,\cite{Banhart1986} and N$^{3-} \approx -1.6\cdot 10^{-10}$\,m$^3$mol$^{-1}$\,\cite{Hohn2009}) and is reflected by the asymmetric error bar for $x = 0.0032$ which is particularly pronounced for $H \perp c$. 
A more sophisticated estimate of the Li$_3$N contribution would require a magnetization measurement on very pure, single crystalline Li$_3$N (of sufficient size due to the small value) or a profound calculation including the effect of the binding electrons. 

Measurements for $H \perp c$ are challenging since smallest misalignments cause significant contributions from $H \parallel c$ due to the giant anisotropy. 
Therefore, these measurements have been performed only for two Fe concentrations so far ($x = 0.0032$ and $x = 0.28$).
A flat glass sample holder was used to accurately mount the single crystals for $H \perp c$.

\subsection*{Measurement protocols for relaxation measurements}\label{protocol}
For time-dependent measurements, the temperature was always stabilized for at least minutes before the magnetic field was changed. Measurement protocols for the relaxation measurements are as follows:
\\
- Relaxation after switching on the field (main article Figs.\,4a-c):
The sample was cooled to low temperatures in zero magnetic field and a first measurement was done confirming $M(H=0) \approx 0$.
After that the magnetic field was ramped to $\mu_0H = 1$\,T (Fig.\,\ref{relax}b) or $\mu_0H = 2$\,T (Fig.\,\ref{relax}c).
Once the field is stable (marked as $t = 0$), $M(t)$ was measured for 4000\,s. 

- Relaxation after switching off the field (main article Figs.\,4d-f):
A magnetic field of $\mu_0H = 7$\,T is applied at a temperature well above the appearance of any detectable hysteresis, here we went to $T = 150$\,K. 
The sample was cooled to $T = 2$ to $16$\,K.
After that, the magnetic field was ramped to zero with a rate of 10\,mT/s.
Once the field is zero (marked as $t = 0$), $M(t)$ was measured for 4000\,s. 
The relaxation process in nominal zero field and small negative fields shown in the main article Figs.\,\ref{relax}e,h can be affected by the finite remanent field of the 9\,T magnet which was empirically found to lie between $\pm 0.1$ and $\pm 1$\,mT.

\subsection*{Details of the fitting procedure}\label{si-fitting}
The values of the relaxation time $\tau$  (main article Fig.\,\ref{relax}h) have been obtained as follows: 
$M_0$ is the magnetization measured directly after the applied field reached zero (which also marks $t = 0$). Accordingly, there are only two free parameters ($\tau$ and $\beta$) for the fit of $M(t) = M_0\,{\rm exp}\left[-(t/\tau)^\beta\right]$.
The exponent $\beta$ accounts for a change of the relaxation rate during time with a faster relaxation at early times and is well established for the analysis of $M(t)$ in SMMs\,\cite{Sessoli2003}.
The obtained $\tau$ and $\beta$ values are given in Supplementary Figs.\,\ref{SI-tau}a,b.
$\beta$ increases from $\approx 0.4$ at $T = 2$\,K to $\beta \approx 0.8$ at $T = 16$\,K (Supplementary Fig.\,\ref{SI-tau}b) and approaches one single relaxation time at high $T$ ($\beta = 1$). A similar increase of $\beta$ with increasing temperature was observed in SMMs\,\cite{Sessoli1997}.
Supplementary Figs.\,\ref{SI-tau}a,b also show the obtained values of $\tau$ and $\beta$ for the relaxation after switching on the applied field. The data was fitted to the stretched exponential $M(t) = M_{\rm equ}\,\left(1-{\rm exp}\left[-(t/\tau)^\beta\right]\right)$. $M_{\rm equ}$ (equilibrium magnetization), $\tau$, and $\beta$ were free parameters. For the fit, $t = 0$ marks the beginning of ramping the field from 0 to $\mu_0H = 1$\,T and $\mu_0H = 2$\,T, respectively.
Accordingly, the applied field is not stable for the first part of the measurement and this introduces a certain error to the extracted values. However, the time to reach a stable field of $\mu_0H = 1$\,T and $\mu_0H = 2$\,T is 200\,s and 250\,s, respectively, which is small compared to the total measurement time of 4000\,s.
Furthermore, we want to emphasize that the temperature-independence of the relaxation in $\mu_0H = 2$\,T below $T = 10$\,K is directly visible from the $M(t)$ data. The roughly constant relaxation time [plotted as ln($\tau$) versus the inverse temperature, red squares in Supplementary Fig.\,\ref{SI-tau}a] found for $T < 10$\,K is in good agreement with this observation.

\subsection*{\textit{M-H} loops for \textit{x} = 0.014, 0.059, and 0.13}\label{loops}
Temperature and sweep-rate dependent $M$-$H$ loops for $x = 0.014, 0.059$ and 0.13 are shown in Supplementary Fig.\,\ref{SI-mag}. 
The step-sizes decrease with increasing $x$. Furthermore, the steps are suppressed more rapidly with increasing temperature when compared to the low Fe concentration $x = 0.0032$.
\\
For each of the $x$ values an applied field of $\mu_0H = 9$\,T is not sufficient to fully reverse the magnetization at $T = 2$\,K and the $M$-$H$ loops are asymmetric and not closed even for the lowest sweep-rate of d$(\mu_0H)$/d$t = 1.1$\,mT/s (right panels in Supplementary Fig.\,\ref{SI-mag}). 
Accordingly, the samples could not be saturated in a zero-field-cooled measurement with similar sweep-rate.
The sweep-rate dependence for $x = 0.13$ is weaker than that observed for lower concentrations (e.g., by comparing the relative change of $H_{\rm c}$ with sweep-rates between the different Fe concentrations).
This indicates an approach to static magnetic ordering in more dense systems where the interaction between the Fe-magnetic moments could lead to the formation of a true ferromagnetic state. 

\subsection*{Correlation of relaxation time and step-position in \textit{M-H} loops}

Within the framework of resonant tunnelling, the relaxation process in the presence of an applied field is expected to be accelerated at the resonance fields (step-positions in $M$-$H$ loops).
This has been experimentally verified for SMMs where $\tau(H)$ shows local minima at the resonance fields\,\cite{Thomas1996, Hernandez1997}.
For Li$_2$(Li$_{1-x}$Fe$_x$)N we found clear anomalies of the relaxation at the resonance fields as shown in Supplementary Fig\,\ref{SI-relax}.
The time-dependent magnetization increases step-like for increasing the field from $\mu_0H = 0.14$ to $0.15$\,T and from $\mu_0H = 0.4$ to $0.5$\,T. 
This increase takes place at the resonance fields as revealed by the steps in the \textit{M}-\textit{H} loops.
However, no clear minimum is observed in $\tau(H)$ for $\tau$ determined from a fit of the $M(t)$ data to a stretched exponential function.
The origin for this deviation from SMM-like behaviour lies, presumably, in the dominance of the relaxation process associated with the large step at $H = 0$.
The smaller steps at $\mu_0H = 0.15$ and 0.55\,T depend much more weakly on the sweep-rate which indicates smaller relaxation times.
When the applied field approaches the first resonance field at $\mu_0H = 0.15$\,T the equilibrium state associated with $0 < \mu_0H < 0.15$\,T has not been reached. Accordingly, the applied field affects both processes, i.e., increasing the field from $\mu_0H = 0.15$ to $0.16$\,T destroys the resonance condition for the small step but enhances the relaxation associated with the large barrier at $H = 0$. Both effects cancel out and $M(t)$ is almost identical in both fields (Supplementary Fig.\,\ref{SI-relax}a).
The same argument explains the large difference between $M(t)$ in $\mu_0H = 0.4$\, and 0.5\,T whereas the difference between $\mu_0H = 0.5$\, and 0.6\,T is small (Supplementary Fig.\,\ref{SI-relax}b). Here the accelerating effect on the relaxation associated with the large barrier at $H = 0$ dominates over the detrimental effect on the tunnelling. 

\end{document}